\pdfoutput=1
\documentclass[twocolumn]{aastex61}

\usepackage{multirow}
\received{...}
\revised{...}
\accepted{...}

\begin{document}

\title{Optical dimming of RW Aur associated with an iron rich corona and exceptionally high absorbing column density}

\correspondingauthor{Hans Moritz G\"unther}
\email{hgunther@mit.edu}

\author[0000-0003-4243-2840]{Hans Moritz G\"unther}
\affil{MIT, Kavli Institute for Astrophysics and Space Research, 77 Massachusetts Avenue, Cambridge, MA 02139, USA}

\author{T. Birnstiel}
\affil{University Observatory, Faculty of Physics, Ludwig-Maximilians-Universit\"at M\"unchen, Scheinerstr. 1,
81679 Munich, Germany}

\author[0000-0002-3860-6230]{D. P. Huenemoerder}
\affil{MIT, Kavli Institute for Astrophysics and Space Research, 77 Massachusetts Avenue, Cambridge, MA 02139, USA}

\author{D. A. Principe}
\affil{MIT, Kavli Institute for Astrophysics and Space Research, 77 Massachusetts Avenue, Cambridge, MA 02139, USA}

\author{P. C. Schneider}
\affil{Hamburger Sternwarte, Universit\"at Hamburg, Gojenbergsweg 112, 21029, Hamburg, Germany}

\author[0000-0002-0826-9261]{S. J. Wolk}
\affil{Harvard-Smithsonian Center for Astrophysics, 60 Garden Street, Cambridge, MA 02138, USA}

\author{Franky Dubois}
\affil{Astrolab IRIS, Ieper, Belgium}
\affil{Vereniging voor Sterrenkunde, Werkgroep Veranderlijke Sterren, Belgium}

\author{Ludwig Logie}
\affil{Astrolab IRIS, Ieper, Belgium}
\affil{Vereniging voor Sterrenkunde, Werkgroep Veranderlijke Sterren, Belgium}

\author{Steve Rau}
 \affil{Astrolab IRIS, Ieper, Belgium}
\affil{Vereniging voor Sterrenkunde, Werkgroep Veranderlijke Sterren, Belgium}

\author{Sigfried Vanaverbeke}
\affil{Astrolab IRIS, Ieper, Belgium}
\affil{Vereniging voor Sterrenkunde, Werkgroep Veranderlijke Sterren, Belgium}
\affil{Center for Plasma Astrophysics, University of Leuven, Belgium}

\begin{abstract}
RW~Aur is a binary system composed of two young, low-mass stars. The primary, RW~Aur~A, has undergone visual dimming events ($\Delta V =2-3$~mag) in 2011, 2014-16, and 2017-2018. Visual and IR observations indicate a gray absorber that moved into the line-of-sight. This dimming is also associated with changes in the outflow. In 2017, when the optical brightness was almost 2~mag below the long-term average we triggered a \emph{Chandra} observation to measure the absorbing column density $N_\mathrm{H}$ and to constrain dust properties and the gas-to-dust ratio of the absorber. In 2017, the X-ray spectrum is more absorbed than it was in the optically bright state ($N_\mathrm{H} = (4\pm 1) \times 10^{23}\;\mathrm{cm}^{-2}$) and shows significantly more hot plasma than in X-ray observations taken before. Also, a new emission feature at $6.63\pm0.02$~keV (statistic) $\pm0.02$~keV (systematic) appeared indicating an Fe abundance an order of magnitude above Solar, in contrast with previous sub-Solar Fe abundance measurements.
Comparing X-ray absorbing column density $N_\mathrm{H}$ and optical extinction $A_V$, we find that either the gas-to-dust ratio in the absorber is orders of magnitude higher than in the ISM or the absorber has undergone significant dust evolution. Given the high column density coupled with changes in the X-ray spectral shape, this absorber is probably located in the inner disk. We speculate that a break-up of planetesimals or a terrestrial planet could supply large grains causing gray absorption; some of these grains would be accreted and enrich the stellar corona with iron which could explain the inferred high abundance.

\end{abstract}


\section{Introduction}
\label{sect:introduction}

The formation of stars and planetary systems from large-scale molecular clouds is a complicated process with many interacting components. After the initial collapse of the cloud, an accretion disk forms around the central proto-star.  Low mass stars ($<3\;M_{\odot}$) at this stage are called classical T Tauri stars (CTTS).
Their circumstellar disks are the sites of planet formation. Disks regulate the angular momentum of the star, and they may launch a disk wind. Typical disks contain a mass of $10^{-3}$ to $10^{-1}\,M_\odot$
and disperse within a few Myrs \citep[see review by][]{2014prpl.conf..475A}, which also sets the 
time-scale for planet formation. During their evolution disks undergo large structural changes. In particular, grains grow to larger sizes and settle in the disk mid-plane leaving
a gas rich disk atmosphere behind \citep[see review by][]{2015PASP..127..961A}.

Gas comprises the majority of the mass in a protoplanetary disk and thus 
controls essential transport processes within the disk such as
angular momentum redistribution and dust grain motion.
For example, gas affects grain growth through the coupling of
gas and dust dynamics \citep{1977MNRAS.180...57W,2001ApJ...557..990T}
as well as the thermal and chemical balance of the disk 
\citep[e.g.][]{2009A&A...501..383W}. 
Images of T~Tauri star disks often have asymmetries or gaps \citep{2013Sci...340.1199V,2013Natur.493..191C,2015ApJ...808L...3A,2015MNRAS.453.1768P,2016ApJ...820L..40A}. Thus, the distribution of grain sizes in the stellar environment differs significantly for different sight lines.

Stellar UV and X-ray emission ionize the upper layers of the disk and in particular the inner disk edge. These ions couple effectively to the star-disk magnetic field. On the disk surface, ions can be accelerated magneto-centrifugally into a disk wind, possibly dredging along dust particles from the disk. Outflows can be clumpy, presumably changing in response to the magnetic field or the stellar irradiation \citep{2012ApJ...758..100B,2014A&A...563A..87E}. The winds remove angular momentum and allow accretion to proceed. From the inner disk edge, gas is funneled onto the star. 

Recent space-based monitoring campaigns with \emph{COROT}, \emph{K2}, and \emph{Spitzer} revealed many different types of variability in the lightcurves of CTTS on time scales as short as hours. Of particular interest are stars with periodic or quasi periodic dips in their lightcurve \citep{2014AJ....148..122G,2015AJ....149..130S}. A well known star of this type is \object{AA Tau} \citep{1999A&A...349..619B}. The inner disk of AA~Tau is seen very close to edge-on and the best explanation is that an asymmetric feature, such as a warp caused by a planet embedded in the disk, moves through our line-of-sight periodically. In addition, AA~Tau has shown a multi-year long dimming event where the visual extinction increased by 4 magnitudes \citep{2013A&A...557A..77B} along with an increased X-ray absorbing column density \citep{2015A&A...584A..51S}. The duration of the event and the variable line emission originating from the upper layers of the disk indicate a position in the disk at a radius of few AU. This absorber has an ISM-like $N_\mathrm{H}/A_V$ ratio.

In this paper we report on new observations of \object{RW Aur} which is a binary system composed of two K-type stars with masses of 1.4 and 0.9~$M_\sun$ and an age close to 10~Myr \citep{1997ApJ...490..353G, 2001A&A...376..982W}. The binary components are separated by 1.4\arcsec{} \citep[semi-major axis 200~au, period 1000~years;][]{2017arXiv170208583C} and are located at a distance of 140~pc \citep{2007A&A...474..653V}. In this paper we concentrate on the primary star which is one of only a few sources with an X-ray detected jet \citep{2014ApJ...788..101S}, indicating outflows in excess of 400~km~s$^{-1}$. The disk mass around RW~Aur~A is $\sim0.001\;M_\sun$ \citep{2005ApJ...631.1134A} and the disk has an intermediate inclination with estimates varying between $45-60^\circ$ \citep{2006A&A...452..897C,2018arXiv180409190R} and $77^\circ$ \citep{2013ApJ...766...12M}, possibly because the inner disk is warped \citep{2016MNRAS.463.4459B}.

In a well-sampled lightcurve reaching back to about 1900, RW~Aur~AB has shown long-term variability and a short dimming event, e.g.\ a one month long dimming in December 1937, every few decades \citep{2017AstBu..72..277B,2018arXiv180409190R}, but it has shown multiple optical dimming events since 2011. The 2011 event lasted about half a year \citep{2013AJ....146..112R} with dimming of $\Delta m_V=2$~mag. Another dimming event started in mid-2014 with $\Delta m_V=3$~mag compared to its typical bright state around $m_V=11$~mag, which has been stable for decades \citep{2013AJ....146..112R}. The stellar flux reached the bright state again in November and December of 2016 before plunging into a new dimming phase. \citet{2017ASPC..510..356L} see indications that the egress from a dim state happens earlier in the IR than in the optical. \citet{2016ApJ...820..139T} find that veiling, a measure of how strongly the continuum from the accretion shock contributes to the optical flux, is stronger in the 2015/16 dimming of RW~Aur than in the bright state; X-shooter data \citep{2016A&A...596A..38F} confirm this finding. On the other hand, \citet{2015A&A...577A..73P} infer no changes in the accretion close to the star, but see increased wind signatures. At the same time, photometry indicates that the 2015 extinction in RW Aur A is gray up to at least $K$ band  \citep{2015A&A...584L...9S,2015IBVS.6126....1A}, while a selective (reddening) $A_V = 0.44$~mag is seen in the bright state \citep{2015IBVS.6126....1A}.

\citet{2006A&A...452..897C} observed a stream of gas on the outer edge of the disk of RW~Aur~A 
and \citet{2013AJ....146..112R} suggested the dimming event in 2011 to be caused by this stream passing the line-of-sight to RW~Aur~A. In ALMA observations \citet{2018arXiv180409190R} confirm the detection by \citet{2006A&A...452..897C} and identify multiple further streams.
Modeling by \citet{2015MNRAS.449.1996D} showed that tidal interaction in the passage of its binary companion, RW~Aur~B, can cause such a stream to form.
However, in the new 2016 dimming event, multiple signatures show variability arising from material close to the star instead. \citet{2015IBVS.6143....1S} report that both increased hot dust emission and optical spectroscopic signatures indicated that there was an increase in absorption by an outflow during the dim state in 2015 \citep{2015A&A...577A..73P,2016A&A...596A..38F,2016MNRAS.463.4459B}.
Coupled with the fact that mass accretion is more stable during this dim state \citep{2016ApJ...820..139T}, it appears that the recent dimming events can only be explained by phenomena close to the star.

RW~Aur~AB has been observed in X-rays several times. \citet{2010A&A...519A.113G} present unresolved \emph{XMM-Newton} data, but \emph{Chandra} is required to resolve the two stellar components. The first \emph{Chandra} observation was taken in 2013 \citep{2014ApJ...788..101S} when RW~Aur~A was in an optically  bright state. \citet{2015A&A...584L...9S} obtained a second dataset in 2015, catching RW~Aur in an optically fainter state. Here, we report on a third dataset observed in 2017, again in an optically dim state. RW~Aur~B is also a variable X-ray source, most likely due to coronal flares.

We first describe observations and data reduction in section~\ref{sect:obs} before we derive results in section~\ref{sect:results}. We discuss the results in section~\ref{sect:discussion} and go through several physical scenarios that might cause the observed signatures in section~\ref{sect:scenarios}. We end with a short summary in section~\ref{sect:summary}.

\section{Observations and data reduction}
\label{sect:observationsanddatareduction}

\label{sect:obs}

We present new and archival X-ray data from \emph{Chandra} and archival data from \emph{XMM-Newton} as well as optical
data from \emph{Chandra}'s aspect camera (ACA) and AAVSO long-term monitoring lightcurves extended into 2017. Details of the X-ray observations are listed in table~\ref{tab:obslog}.

\subsection{\emph{XMM-Newton} X-ray data}
\label{sect:emphxmmnewtonxraydata}
RW~Aur~AB was centered in the field of view in the \emph{XMM-Newton} observation, but the count rate in the RGS is too low to analyze the grating spectrum. We reduced the data with SAS version 15.0.0 following the standard procedures to screen periods of high-background flux. We extracted a source region centered on the unresolved binary (AB) with a radius of 30\arcsec{} in the PN and MOS detectors. We selected source-free background regions on the same chip (avoiding the chip edges where noise can be higher) with a radius of 50\arcsec{} in the PN and 105\arcsec{} for the MOS detectors.

\subsection{\emph{Chandra} X-ray data}
\label{sect:emphchandraxraydata}

Observations were taken roughly two years apart, where the last observation is split into two orbits separated by two days. All processing is done with CIAO version 4.9 and CALDB 4.7.1. RW~Aur~B, the brighter member of the binary in X-rays, was bright enough to be at the risk of pile-up in the 2017 observations but at the fluxes observed the pile-up fraction is less than 5\% even in the brightest pixel; previous data were taken in a 1/8 sub-array mode that reduces the field-of-view and the frame time to alleviate this problem. For RW~Aur~A, the focus of this paper, pile-up is not significant in any of the observations.

\begin{figure*}[ht!]
\plotone{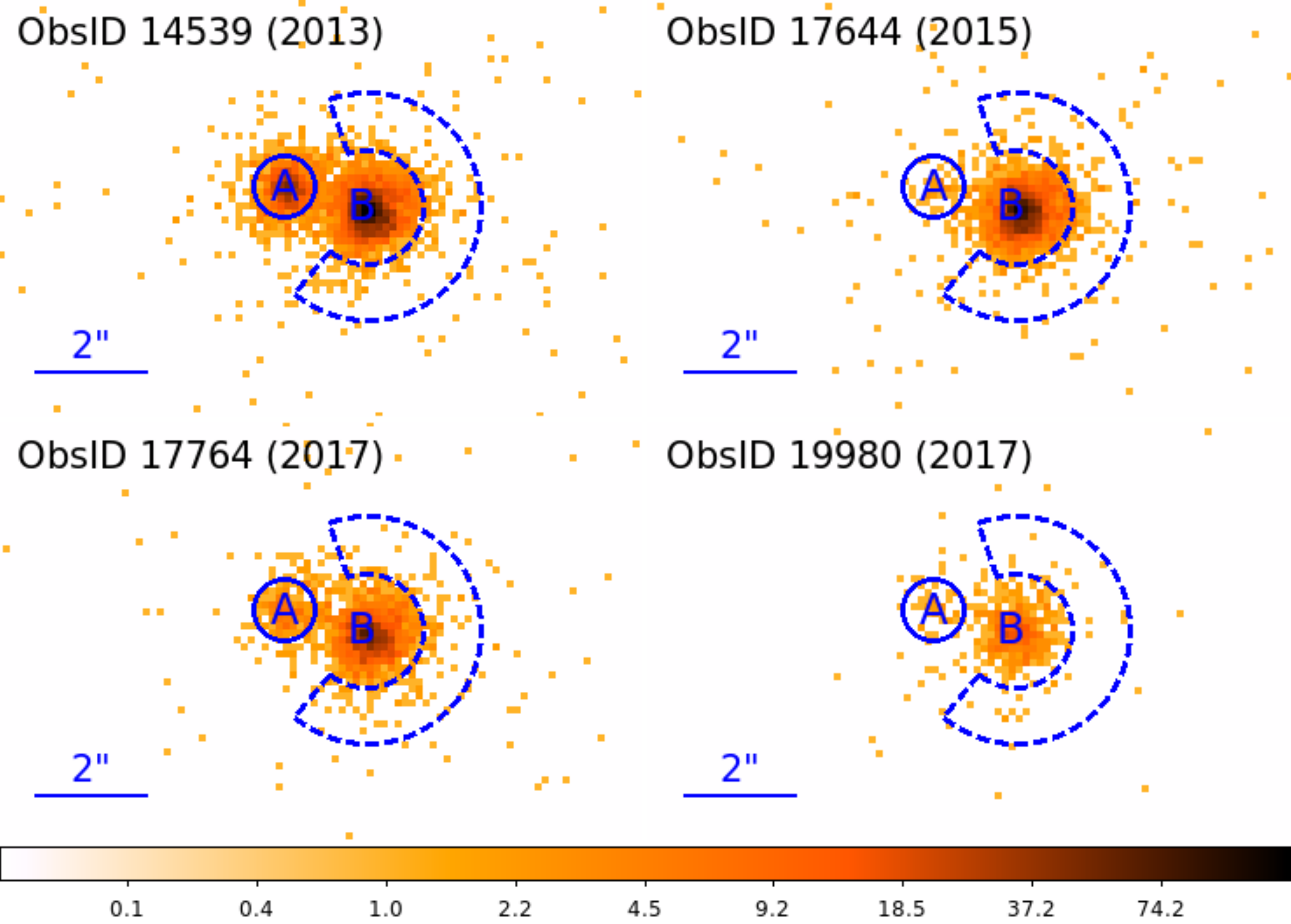}
\caption{Images of RW~Aur~A and B in all four \emph{Chandra} observations. The source region for RW~Aur~A (solid line) and annulus (dashed line) designed to capture the contamination by the wings of the PSF of the brighter RW~Aur~B are marked. The color scale shows counts per spatial bin. (The spatial bin size, 0.123~arcsec, in this image is smaller than the physical size of the ACIS pixels.) Note that the exposure time differs between observations and thus the same number of counts does not represent the same flux.}
    \label{fig:regions}
\end{figure*}
We limit our analysis to the energy band 0.3-9.0~keV. Apertures with radius 0.54~arcsec (covering 75\% of the point-spread function - PSF) were used to extract spectra from RW~Aur A and B. We measure the background flux from a large, source-free region on the same detector and find that the expected background flux in each source region is $<0.2$~counts in any one observation. 
The B component of the RW~Aur system is much brighter in X-rays than the A component. Therefore, the wings of the PSF from RW~Aur~B contribute to the data extracted for RW~Aur~A, while the reverse contamination is not relevant. 
Thus, we define an annulus centered on RW~Aur~B with an inner and outer radius of 1 and 2~arcsec, respectively, like it was done in \citet{2015A&A...584L...9S}. This corresponds to the radii covered by the extraction region of RW~Aur~A. We remove a segment of $\pm60^\circ$ around the position of RW~Aur~A from this annulus and use the remaining area (figure~\ref{fig:regions}) to estimate the number of counts due to contamination by RW~Aur~B (column ``contam'' in table~\ref{tab:obslog}). This is a small fraction of the counts from RW~Aur~A except for ObsID~17644 (2015).

\begin{table*}
\caption{\label{tab:obslog} X-ray observations}
\begin{tabular}{ccccccccc}
\hline \hline
Observatory & ObsID & Date & MJD & Exp. time$^a$ & livetime$^b$ & Mode & RW Aur A & contam \\
 &  &  &  & $\mathrm{ks}$ & $\mathrm{ks}$ &  & counts & counts \\
\hline
XMM-Newton & 0401870301 & 2007-02-21 & 54123.5 & 36.4 & 34.5 & full frame & \multicolumn{2}{c}{unresolved} \\
Chandra/ACIS-S & \dataset[14539]{ADS/Sa.CXO\#obs/14539} & 2013-01-12 & 56304.1 & 60.9 & 54.5 & 1/8 subarray & 801 & 22.3 \\
Chandra/ACIS-S & \dataset[17644]{ADS/Sa.CXO\#obs/17644} & 2015-04-16 & 57128.3 & 40.2 & 35.1 & 1/8 subarray & 44 & 20.1 \\
Chandra/ACIS-S & \dataset[17764]{ADS/Sa.CXO\#obs/17764} & 2017-01-09 & 57762.3 & 41.1 & 38.5 & full frame & 173 & 18.2 \\
Chandra/ACIS-S & \dataset[19980]{ADS/Sa.CXO\#obs/19980} & 2017-01-11 & 57764.1 & 14.5 & 10.2 & full frame & 36 & 4.5 \\
\hline
\end{tabular}
\\$^a$: Time for exposure start to exposure end.\\$^b$: Live time in \emph{Chandra}, i.e.\ corrected for the deadtime during readout. In \emph{XMM-Newton} we give the \emph{ONTIME}, the sum of all good time intervals of the PN chip that detected RW~Aur.
\end{table*}

\subsection{\emph{Chandra} optical data}
\label{sect:emphchandraopticaldata}

\emph{Chandra} has a small optical telescope in the aspect control assembly (ACA). This is a CCD detector with a wide bandpass from about 0.4 to 1.1~$\mu$m. The color conversion to standard filters is not calibrated for stars of arbitrary spectral shape. Only a few regions on the CCD are read out and transmitted to the ground. One of these slots was placed on the science target for ObsIDs 17644, 17767, and 19980. The image is intentionally defocussed and RW~Aur is not resolved. Aperture photometry is performed using the CCD noise model, see \citet{2010ApJS..188..473N} for details.

\subsection{AAVSO data}
\label{sect:aavsodata}

We retrieved data for RW~Aur from the database of the American Association of Variable Star Observers (AAVSO) in four bands: Visual and standard $V$, $R$, and $I$ filters. Many observers with different instrumental set-ups contributed to this data collection, but specifically the $B$, $V$, and $R$ data close to the\emph{Chandra} observations in 2017 are taken with a 684~mm aperture Keller F4.1 Newtonian New Multi-Purpose Telescope of the public observatory Astrolab Iris, Zillebeke, Belgium\footnote{www.astrolab.be}. The CCD detector assembly is a Santa Barbara Instrument Group STL 6303E operating at -20$^\circ$~C. 
The 9~$\mu$m physical pixels are read out binned to $3 \times 3$~pixels, which is 1.86~arcsec per pixel. 
The $B$, $V$, and R filters are from Astrodon Photometrics, and have been shown to reproduce the Johnson/Cousins system closely. Differential photometry relative to stars in the field is conducted with the LesvePhotometry reduction package.

\section{Results}
\label{sect:results}

\subsection{Lightcurves}
\label{sect:lightcurves}

We first look at the long-term optical lightcurve and how the timing of the \emph{Chandra} observations relates to the optical dimming events. Then, we turn to the \emph{Chandra} count rates. Last, we present the lightcurve of the \emph{XMM-Newton} observation, which includes a large X-ray flare but lacks dense optical monitoring.

\begin{figure*}[ht!]
    \centering
    \includegraphics[width=0.8\textwidth]{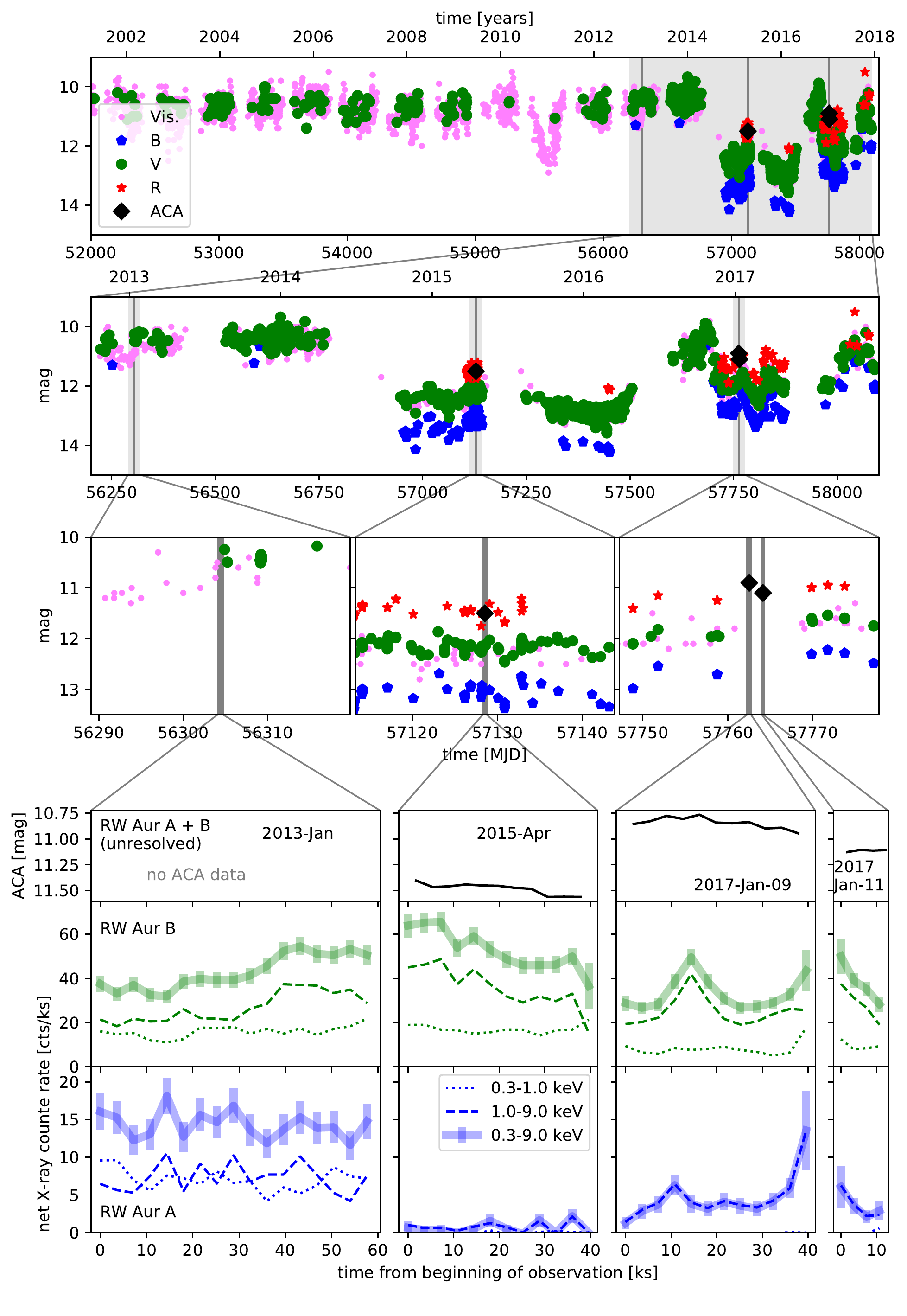}
    \caption{\emph{top three rows} Optical lightcurve of RW~Aur~AB from AAVSO data. The system is not spatially resolved in these observations. Grey vertical bands mark the times of \emph{Chandra} observations.  ``Vis.'' are visual measurements without a standard filter, but the effective bandpass is typically fairly close to $V$ band. ACA is the band of the \emph{Chandra} aspect control assembly.
\emph{bottom block: Chandra} lightcurves from ACA (optical, unresolved), X-rays from RW Aur B (green), and X-rays from RW Aur A (blue). For clarity, error bars for $1\sigma$ uncertainties are only shown for the full band lightcurve. The increasing contamination on \emph{Chandra}/ACIS means that a source with constant flux will produce a lower raw count rate in later observations. There are so few soft X-ray counts in RW~Aur~A in 2015 and 2017 that the dotted line almost exactly falls on the x-axis in 2015 and 2017. All lightcurves are background subtracted and binned to one hour.}
    \label{fig:lc}
\end{figure*}
\subsubsection{Optical lightcurves}
\label{sect:opticallightcurves}

Figure~\ref{fig:lc} (top rows) shows a long-term lightcurve of the RW~Aur~AB system. The first \emph{Chandra} observation (marked by the first gray vertical line) took place during a bright state which had been the long-term average for several decades with $m_V\approx 10.5$~mag. The second \emph{Chandra} dataset was taken in an obscured state with $m_V=12$~mag in 2015. At the end of 2016, RW~Aur~AB briefly reached a bright state again, before fading back to $m_V=11.7$ during the \emph{Chandra} observations in January 2017. Later in 2017, the flux climbed back to a bright state again for a short time, just to drop again towards the end of 2017. Unfortunately, there is a gap in the optical data of a few days right around the \emph{Chandra} observation. However, while RW~Aur~AB is known to have some variation on time scales of days and hours \citep{2016MNRAS.463.4459B}, the lightcurve around the 2017 observations seems relatively smooth.
The \emph{Chandra}/ACA monitoring shows steady lightcurves with smooth variability on the 0.1~mag level during the observations and a difference of about 0.2~mag between the two observations in 2017 which are about two days apart. \citet{2016MNRAS.463.4459B} determine that the variability of RW~Aur~A on time scales of hours is irregular, which can be explained by changes in the accretion rate. Our data are consistent with this scenario, but the \emph{Chandra} observations by itself are too short to distinguish this from the periodic variability that hot (accretion) or cold (magnetic) spots would produce when they rotate in and out of view.

We interpolate the optical lightcurves to obtain values during the \emph{Chandra} observation (table~\ref{tab:BVR}). In 2015, there are observations within a few hours of the \emph{Chandra} data in all bands. In all cases, the uncertainty is dominated by the variability of the target and we estimate the error on the optical magnitudes to 0.2~mag. Within this uncertainty, RW~Aur~AB is brighter by 0.5~mag in $B$, $V$, $R$, and the ACA band during the 2017 \emph{Chandra} observation, compared with 2015. We do not see significant changes in the optical color between those observations.

None of the optical data presented here resolves the two components of the RW~Aur~AB system. \citet{2015IBVS.6126....1A} show that RW~Aur~B is also variable, but to a much lesser degree than RW~Aur~A. Comparing observations from 1994 and 2014 they find that RW~Aur~B has become brighter by about 0.7~mag with almost no color variability. They suggest this to be due to a dust cloud with large grains moving through our line-of-sight, analogous to what is discussed for RW~Aur~A. We assume that this evolution is slow and subtract the long-term average of the RW~Aur~B fluxes (table~\ref{tab:BVR}) to obtain the flux of RW~Aur~A. Within the uncertainties, we do not see changes in color. On the other hand, changes in $B-V$ color up to about 0.3~mag ($1\sigma$ confidence range) are also possible.

\subsubsection{\emph{Chandra} X-ray  lightcurves}
\label{sect:emphchandraxraylightcurves}

Figure~\ref{fig:lc} (bottom) shows X-ray lightcurves for both components of the RW~Aur system and the ACA lightcurve where data exist. RW~Aur~B shows X-ray variability in every observation, but the average count rates are all similar. During the first observation, the flux increases smoothly by 30\% and decreases by a similar amount in the second observation. In 2017 there is a short flare in the hard band that lasts about 5~ks. The second observation in 2017 shows rapid decline of the hard X-ray flux, possibly the tail end of a flare. Note that the count rate in the soft band is lower for the later observations, because contamination builds up on \emph{Chandra}/ACIS and the effective area declines between epochs. 
RW~Aur~A is fainter than RW~Aur~B in all observations. For RW~Aur~A, significant variability within an observation is seen only towards the end of the first observation in 2017 when the flux triples. The count rate in 2015 is more than an order of magnitude below the value seen in 2013 before the optical dimming started. In the 2017 observations, the count rate in the hard band reaches the pre-dimming level again, but essentially no signal is detected in the soft band.

\begin{table*}
\caption{\label{tab:BVR} Optical brightness for RW~Aur.}
\begin{tabular}{cccccc}
\hline \hline
RW Aur & B$^a$ & AB & AB$^b$ & A$^c$ & A$^c$ \\
year & 2014 & 2015 &  2017 & 2015 & 2017 \\
 unit& mag & mag & mag & mag & mag \\
$B$ & $14.5 \pm 0.3 $ & $13.0 \pm 0.2 $ & $12.5 \pm 0.2 $ & $13.3 \pm 0.3 $ & $12.7 \pm 0.2 $\\
$V$ & $13.2 \pm 0.3 $ & $12.2 \pm 0.2 $ & $11.7 \pm 0.2 $ & $12.8 \pm 0.4 $ & $12.0 \pm 0.3 $\\
$R$ & $12.3 \pm 0.3 $ & $11.5 \pm 0.2 $ & $11.2 \pm 0.2 $ & $12.2 \pm 0.5 $ & $11.7 \pm 0.4 $\\
\hline
\end{tabular}\\
$^a$:\protect{\citet{2015IBVS.6126....1A}}\\
$^b$: Interpolated between observations taken a few days before and after the \emph{Chandra} observations.\\
$^c$: These values are inferred from the unresolved measurements of RW Aur AB by subtracting the RW~Aur~B flux from 2014, assuming that the flux from RW~Aur~B changes only on long time scales. See section~\ref{sect:opticallightcurves} for details.\\
\end{table*}

\subsubsection{\emph{XMM-Newton} lightcurves}
\label{sect:emphxmmnewtonlightcurves}

\begin{figure}[ht!]
    \plotone{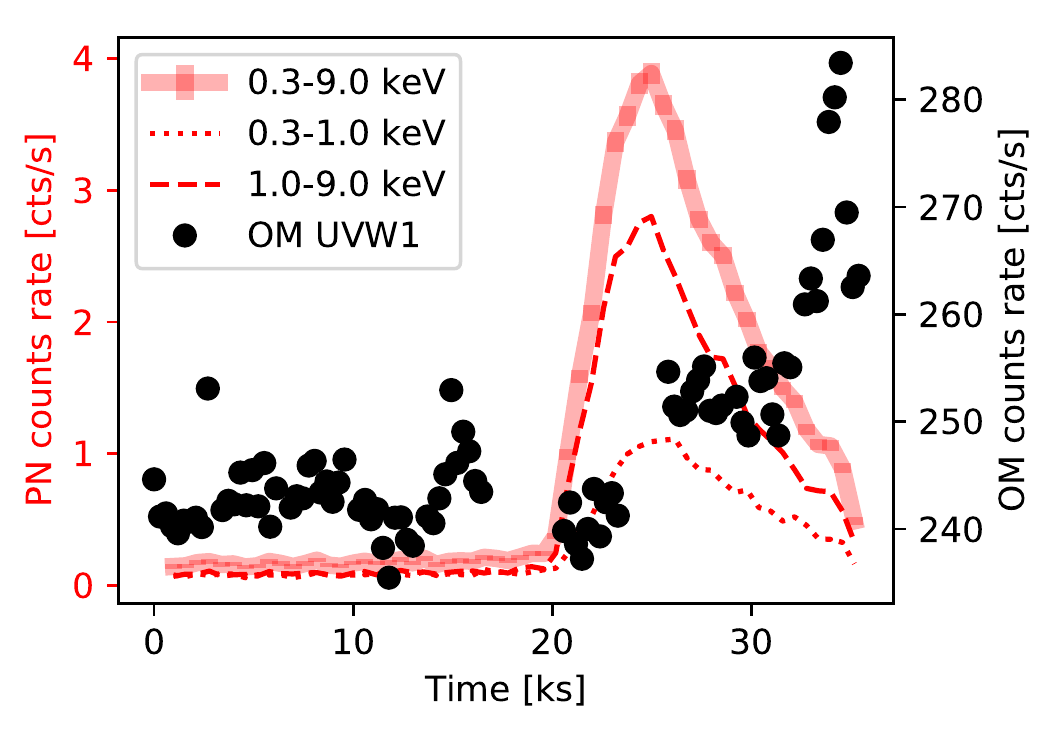}
    \caption{Count rates in the \emph{XMM-Newton} observation in 2007 which did not resolve RW~Aur~AB. The X-ray count rate is mostly flat up to 20~ks, when a massive flare started. As in Fig.~\ref{fig:lc}, error bars are shown only for the full-band lightcurve for clarity; they are comparable to the thickness of the line. In the flare, the count rate increases by factor of 20. The OM observed in a UV band (errors bars are smaller than symbols). There is an increase in OM count rate, but it rises several hours after the X-ray flare.}
    \label{fig:xmmlc}
\end{figure}

The first half of the \emph{XMM-Newton} observation in 2007 shows quiescent emission. A large flare erupted around 20~ks into the observation with an increase in count rate by a factor of $\sim20$ (Fig.~\ref{fig:xmmlc}). At the same time, the hardness ratio ($\frac{H-S}{H+S}$, where $H$ is the count rate in the hard band of Fig.~\ref{fig:xmmlc} and $S$ the count rate in the soft band) increases from -0.4 to 2.5 at the peak of the flare, before it decays back down to 0 at the end of the observation.

We compared the centroid of the X-ray emission in the quiescent and the flare phase and they agree to about 0.1\arcsec{} (with statistical uncertainties of 0.4\arcsec{} and 0.2\arcsec{}, respectively). The absolute astrometry of the \emph{XMM-Newton} observation is insufficient to associate the centroid with either RW~Aur A or B. If RW~Aur~A and B contribute to the quiescent emission, the pre-flare centroid would be located between them close to the brighter component (in the \emph{Chandra} observations RW~Aur~B is always brighter than RW~Aur~A). RW~Aur~A and B are separated by 1.4\arcsec{}. The count rate (and thus centroid of the detected events) is certainly dominated by the flaring source during the flare. So, if RW~Aur A and B had equal X-ray brightness in the quiescent phase, the centroid position in the flare should shift by 0.7\arcsec{}. A much smaller value is observed and we thus conclude that the flare most likely occurs on the same component that dominated the quiescent X-ray emission, but we cannot identify from the \emph{XMM-Newton} observation if this component is RW~Aur~A or B.

\subsection{X-ray spectra}
\label{sect:xrayspectra}

We first discuss the \emph{XMM-Newton} data from 2007 where RW~Aur~AB is unresolved, but we have the most signal (section~\ref{sect:xmmdata}). Next, we turn to the \emph{Chandra} observations where RW~Aur~B (section~\ref{sect:chan:rwb}) and RW~Aur~A (section~\ref{sect:chan:rwa}) are spatially separated, but the spectra from RW~Aur~A could be contaminated by the close-by, brighter RW~Aur~B. Of particular importance is a strong emission feature at 6.63~keV in the spectrum of RW~Aur~A in 2017, which we analyze in detail in section~\ref{sect:2017}.

We use the photospheric abundances given in table~1 of \citet{2009ARA&A..47..481A} as reference throughout this paper. Uncertainties in this section are given as 90\% confidence ranges.

\subsubsection{2007 \emph{XMM-Newton} data}
\label{sect:2007emphxmmnewtondata}
\label{sect:xmmdata}

\begin{figure}[ht!]
\plotone{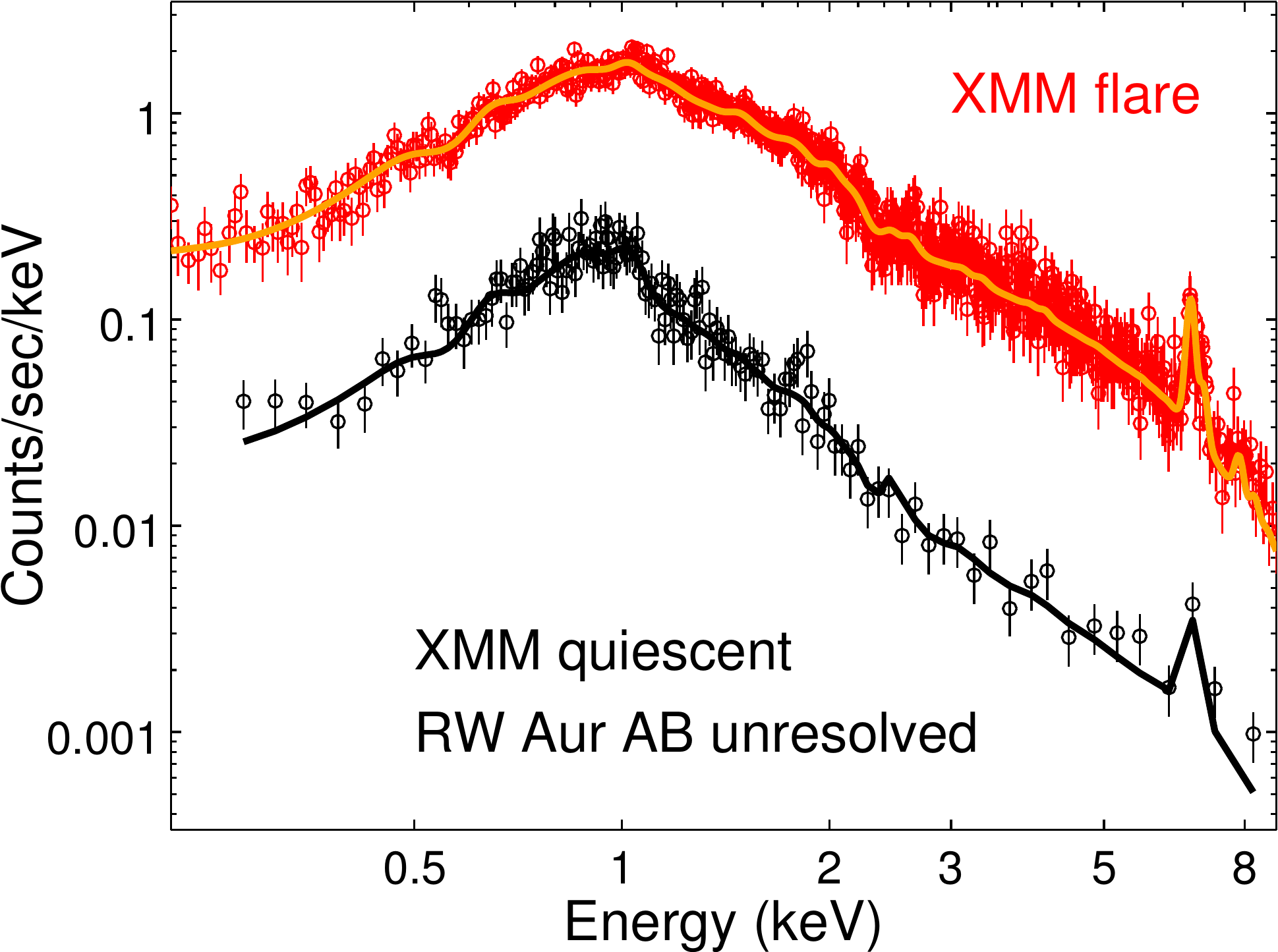}
\caption{\emph{XMM-Newton}/PN data with best-fit models in the quiescent and the flare phase. The RW~Aur~AB binary is unresolved. Data are binned to 20 counts per bin. Error bars in the plot are $1\sigma$ statistical uncertainties. The lines show the best-fit model with the parameters from table~\ref{tab:specfits}.}
    \label{fig:XMM}
\end{figure}

The \emph{XMM-Newton} spectra from RW~Aur~AB are displayed in figure~\ref{fig:XMM} separated into the quiescent phase (the first 20~ks of the observation) and the flare phase (after 20~ks), see figure~\ref{fig:xmmlc}. Our spectral model consists of two optically thin, collisionally excited plasma model components \citep[APEC,][]{2012ApJ...756..128F} and a cold photoelectric absorber. We fit abundances in three groups of elements. We combine Mg, Si, and Fe into one group, because they all have very similar first ionization potential (FIP) values (7.6-8.1 eV). The FIP of Ne is 21.6~eV and is fitted on its own. We fix the abundances of elements with medium FIP values (S, O, N, and C all have FIP between 10 and 15 eV) to 1 because absolute abundances cannot be determined without grating spectroscopy; if we, for example, multiplied all abundances in the model by three and reduced the emission measure  $EM = \int n_i n_e \mathrm{d}V$ ($n_i$ and $n_e$ are the ion and electron number density, respectively, which are integrated over the emitting volume $V$) by the same factor, the model would predict an almost identical spectrum. We investigated a possible change in $N_\mathrm{H}$ or abundance between the quiescent and the flare phase, but did not see statistically significant differences and settled on a model where $N_\mathrm{H}$ and abundances are the same in the quiescent and the flare phase. Furthermore, we coupled the temperature components, so that the fits in the quiescent and the flare phase have the same temperatures, but different emission measures. This simplifies the interpretation, while still providing a good fit. The model thus has nine parameters: $N_\mathrm{H}$, the abundances of Ne and Fe, the temperature of the cool and the hot emission component, and the $EM$ of the cool and hot component in the quiescent and the flare phase. We fit the signal from the PN, MOS1, and MOS2 cameras on \emph{XMM-Newton} simultaneously using a $\chi^2$ statistic. The data are binned to 20 counts per bin. 
Fit results are shown in table~\ref{tab:specfits}. The reduced $\chi^2$ of the model is only 0.7. Simpler models with fewer parameters (e.g.\ only one emission component with a single temperature) can still fit the data with a reduced $\chi^2$ around 1, but they show systematic deviations (e.g.\ the data below 1~keV is consistently underpredicted) and thus we chose to present the two-temperature model.

The 2007 \emph{XMM-Newton} spectrum of RW~Aur AB displays IFIP (inverse first ionization potential) abundances where the abundance of elements with a high FIP such as Ne are enhanced and Fe is reduced compared to solar abundances as seen in the Ne/Fe ratio. In the flare, the $EM$ in the cool component increases by a factor of four, while the $EM$ of the hot component increases by almost a factor of 30.

The observed spectrum shows an emission feature around 6.7~keV (figure~\ref{fig:XMM}) both before and during the flare. This feature is fully compatible with Fe~{\sc xxv} emission and is well-described by the plasma model fits. The hotter component of the plasma in the model is close to the peak formation temperature $\mathrm{k}T=5.5$~keV (k is the Boltzmann constant) of the Fe~{\sc xxv} 6.7~keV line. Given the higher $EM$ in the flare, this feature is stronger there. Observing a 6.7~keV emission line from hot plasma is not surprising, we only note this here because the \emph{Chandra} spectra show a feature at a slightly lower energy in 2017 that is discussed in detail in section~\ref{sect:2017}.


\begin{table*}
\caption{\label{tab:specfits} Parameters of X-ray model fits. Uncertainties are 90\% confidence ranges.}
\begin{tabular}{lccccccccc}
\hline \hline
RW Aur & AB & AB & B & B & B & A & A & \multicolumn{2}{c}{A}\\
year & 2007 & 2007 & 2013 & 2015 & 2017 & 2013 & 2015 & \multicolumn{2}{c}{2017$^{a}$}\\
instrument & XMM & XMM & Chandra & Chandra & Chandra & Chandra & Chandra & \multicolumn{2}{c}{Chandra}\\
           & quiescent & flare & & & & & & 2 temp & 1 temp\\
\hline
$N_\mathrm{H}$ [$10^{21}$~cm$^{-2}$] & \multicolumn{2}{c}{$2.62\pm 0.02$} &
             \multicolumn{3}{c}{$3\pm 1$} &
             $1.1\pm 0.1$ & $70^{+50}_{-30}$ $^c$ & $400 \pm 100^c$ & $300 \pm 100^c$\\
k$T_1$  [keV]& \multicolumn{2}{c}{$0.78\pm 0.04$}
           & \multicolumn{3}{c}{$0.38 \pm 0.05$} &
           $0.6^{+0.1}_{-0.2}$ & =0.6 & $1.4^{+0.6}_{-0.3}$ & $3.1^{+1.8}_{-1.3}$\\
$EM_1$  [$10^{52}$~cm$^{-3}$] & $11\pm 2$ & $45\pm 9$ &
         $4^{+4}_{-2}$ & $6^{+6}_{-3}$ & $3^{+4}_{-2}$ &
         $1.1 \pm 0.4$ & $0.1^{+13}_{-0.1}$ & $50^{+90}_{-20}$ & $17^{+45}_{-9}$\\
k$T_2$  [keV] & \multicolumn{2}{c}{$6.0\pm 0.3$} &
       \multicolumn{3}{c}{$2.0 \pm 0.2$} &
       =20 & =20 & =20 & --\\
$EM_2$  [$10^{52}$~cm$^{-3}$]  & $4.7\pm 0.5$ & $128 \pm 2$ &
      $7 \pm 1$ & $9 \pm 1$ & $8 \pm 1$ &
      $1.0 \pm 0.2$ & $0.5^{+0.8}_{-0.2} $ & $1.9 \pm 1.9$ & --\\
abund$^b$: Fe & \multicolumn{2}{c}{$0.12\pm 0.03$} & 
         \multicolumn{3}{c}{$0.3^{+0.2}_{-0.1}$} &
         $0.5^{+0.2}_{-0.1}$ & =0.5 & $15^{+40}_{-8}$ & $5^{+3}_{-2}$\\
abund$^b$: Ne & \multicolumn{2}{c}{$1.2^{+0.5}_{-0.4}$} & 
         \multicolumn{3}{c}{$2.2^{+1.2}_{-0.8}$} &
         =1 & =1 & =1 & =1\\
\hline
red. $\chi^2$ (dof) & \multicolumn{2}{c}{0.7 (1726)} & 
            \multicolumn{3}{c}{0.7 (251)} &
            \multicolumn{4}{c}{Cash statistic}\\
\hline
observed flux$^d$ & 4.7 & 78 &
             3.0 & 4.5 & 3.3 &
             1.2 & 0.2 & 2.7 & 2.9\\
intrinsic $\log L_X^e$ & 30.3 & 31.4 &
             30.1 & 30.3 & 30.1 &
             29.5 & 29.0 & 31.6 & 30.6\\
\hline
\end{tabular}\\
$^a$: Two different models are fit to the same dataset; the first one (2 temp) with two emission components at different temperature, the second one (1 temp) with a single temperature \\
$^b$: Relative to solar abundances from \protect{\citet{2009ARA&A..47..481A}}\\
$^c$: These numbers should be treated as lower limits. See section~\ref{sect:chan:rwa} for discussion.\\
$^d$: in units of $10^{-13}$erg~s$^{-1}$~cm$^{-2}$ for the energy range 0.3-9.0~keV. Value is given for best fit model without uncertainties.\\
$^e$: in units of erg~s$^{-1}$ for the energy range 0.3-9.0~keV. Value is given for best fit model without uncertainties.
\end{table*}

\subsubsection{\emph{Chandra} spectra from RW Aur B}
\label{sect:emphchandraspectrafromrwaurb}
\label{sect:chan:rwb}

\begin{figure}[ht!]
\plotone{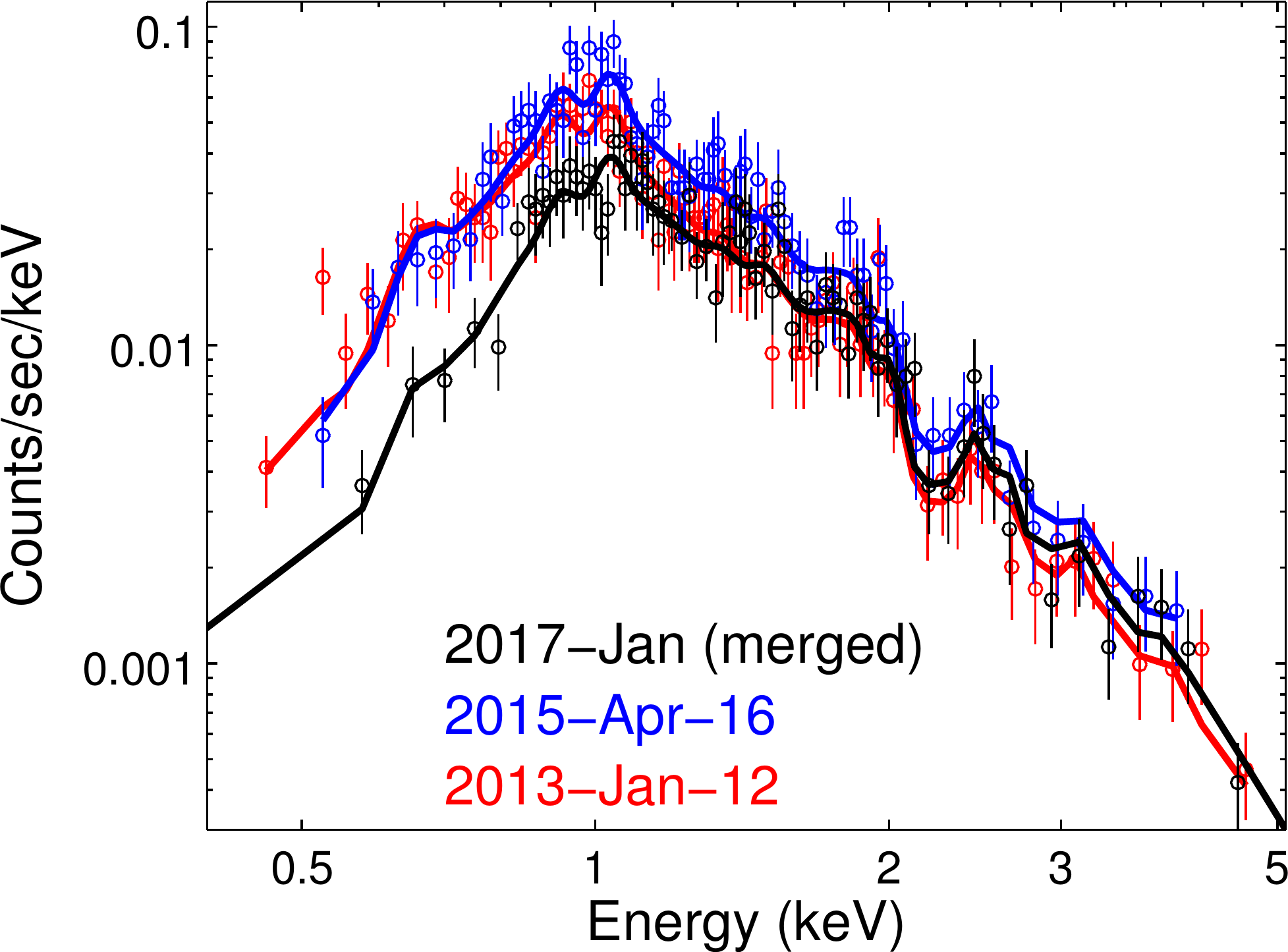}
\caption{RW~Aur~B spectra extracted from \emph{Chandra} data. Error bars in the plot are $1\sigma$ statistical uncertainties. The lines show the best-fit model with the parameters from table~\ref{tab:specfits}. The spectra are very similar, except for the lower number of counts at low energies in the 2017 observation. This is due to increased contamination on the ACIS detector and taken into account in fitting.}
    \label{fig:RWAurB}
\end{figure}
The three \emph{Chandra} datasets from RW~Aur~B are qualitatively similar to each other (figure~\ref{fig:RWAurB}); differences in the \emph{Chandra} effective area at low-energies are taken into account by using the appropriate ARF file in the fitting. Background is negligible in all cases. The two datasets from 2017 are treated separately in the fit, but we impose the condition that the fit parameters are identical for those two datsets (the datasets are only merged for display purposes in figure~\ref{fig:RWAurB}).
We fit the same model of one absorption component and two APEC plasma models that we used for the \emph{XMM-Newton} data. Again, the value of the reduced $\chi^2$ is low, but if we reduce the number of model parameters (e.g.\ fit only one APEC component or fix the Ne and Fe abundance at solar values) we see systematic residuals in the fit.

Compared to the \emph{XMM-Newton} data with its prominent flare, the plasma temperatures are much lower. There is no plasma as hot as the component that gives rise to the 6.7~keV line in the \emph{XMM-Newton} data. The absorbing column density and the abundance ratios are similar to the values found from the \emph{XMM-Newton data}. It is thus plausible that the \emph{XMM-Newton} observation was dominated by emission from RW~Aur~B.

\subsubsection{\emph{Chandra} spectra from RW~Aur~A}
\label{sect:emphchandraspectrafromrwaura}
\label{sect:chan:rwa}
\begin{figure}[ht!]
    \plotone{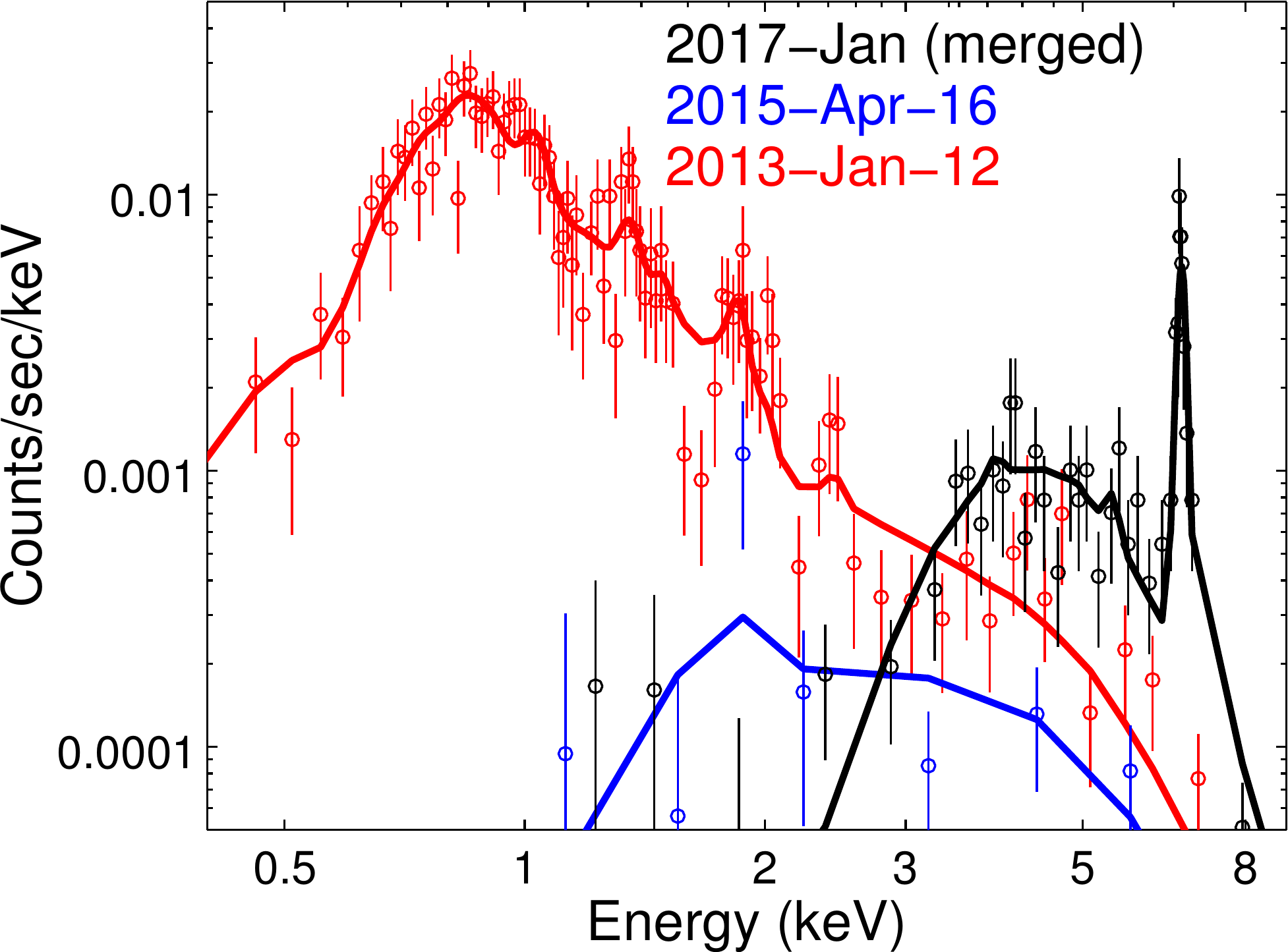}
    \caption{X-ray spectra for RW~Aur~A. Points are observational data and lines show models from table~\ref{tab:specfits} (for 2017, the "2 temp" model is shown). The two observations in 2017 are combined for display purposes only, and all data are shown background-subtracted for display purposes. Error bars indicate $1\sigma$ uncertainties.}
    \label{fig:spec:rwa}
\end{figure}

\emph{Chandra} spectra from RW~Aur~A for the epochs 2013, 2015, and 2017 are shown in figure~\ref{fig:spec:rwa}.
In 2017, RW~Aur~A is significantly brighter at high energies and shows a strong emission feature located at 6.63~keV (inset in figure~\ref{fig:spec_17} and discussed in detail in section~\ref{sect:2017}) that is not seen in previous observations. At the same time, there is very little signal at soft energies after accounting for the contamination by the wings of the PSF from RW~Aur~B. 
We again fit a model with two thermal emission components and one absorption component. We add a model for the contamination by the RW~Aur~B PSF with the parameters of RW~Aur~B in table~\ref{tab:specfits} to each RW~Aur~A spectral fit. The normalization of these components is set to account for the small fraction of the RW~Aur~B PSF that falls into the RW~Aur~A spectral extraction region. We calculate a normalization factor from the counts observed in the dashed regions in figure~\ref{fig:regions} and then multiply it by the ratio of the areas of the RW~Aur~A extraction region and the dashed region. Since the total number of counts in the RW~Aur~A spectra is significantly lower than in the \emph{XMM-Newton} data or in the RW~Aur~B spectra, we perform the fit using the Cash statistic (statistic \texttt{cash} in Sherpa, see \citet{2007ASPC..376..543D} and \citet{1979ApJ...228..939C}), which correctly accounts for the Poissonian likelihood in low-count bins, but has the downside that is does not provide a goodness-of-fits statistic such as the reduced $\chi^2$. As in section~\ref{sect:chan:rwb} the datasets from 2017 are kept separate in the fitting process and are combined for display purposes only.

The decrease in soft ($\lesssim$ 2 keV) X-ray flux from RW Aur~A is the result of a
variable absorbing column density whose extinction is higher in 2015 and 2017 than in the previously
observed optically bright state (2013; red in figure~\ref{fig:spec:rwa}); the contamination of ACIS has a much smaller effect (compare with figure~\ref{fig:RWAurB}).

Like \citet{2014ApJ...788..101S} we find that a hot plasma component is required to fit the RW~Aur~A spectrum in 2013, but the temperature of this component is poorly constrained. \citet{2014ApJ...788..101S} discuss this in detail and show that a powerlaw component actually gives a slightly better fit than a collisionally-excited plasma, but we do not know of a physical process that could produce powerlaw emission in RW~Aur~A. Following \citet{2014ApJ...788..101S} we fix the temperature of the hotter component at 20~keV. Furthermore, we find that the data constrain the Ne abundance insufficiently, and we thus fix this value at 1. 
Given the low signal in 2015, even more assumptions are needed to fit a model. We fix both temperatures and the Fe abundance at the numbers found for the 2013 data and leave only the absorbing column density $N_\mathrm{H}$ and the normalization of the two emitting components for the fit. The fitted value for $N_\mathrm{H}$ depends on these assumptions. For example, a cooler $\textnormal{k}T_1$ would predict more emission at lower energies and thus require a larger $N_\mathrm{H}$ to explain the absence of signal below 1~keV. However, the fact that the observed flux dropped by two orders of magnitude or more below 1~keV compared to the data from 2013 while the drop is less severe at higher energies suggests a strongly increased $N_\mathrm{H}$ \citep[for a detailed discussion of the 2015 spectrum see][]{2015A&A...584L...9S}.

In 2017, the spectrum is very different again (figure~\ref{fig:spec:rwa}). There is very little signal below 2~keV, while the flux $>3$~keV is several times stronger than in 2013 or 2015. In particular, there is a very strong emission feature between 6 and 7~keV, which is not seen in the previous \emph{Chandra} data. This feature is discussed in detail in the next section. In table~\ref{tab:specfits} we show two different models for the 2017 data. First, we fit a model similar to the models for 2013 and 2015 with a hot plasma component at 20~keV; this hot component has an $EM$ comparable to the other datasets; however the temperature of the cooler component is about twice the value seen in 2013, and the cool emission measure is about 50 times higher. Notably, an Fe abundance 15 times higher than in the sun is required to match the emission feature between 6 and 7~keV. Such a large emission measure will also produce copious flux below 2~keV and a very high absorbing column density of $400 \times 10^{21}$~cm$^{-2}$ is required to explain why we do not observe any flux in that region. Alternatively, we fit a model with just a single emission component at $\mathrm{k}T\approx3$~keV. Due to the higher temperature, the Fe emissivity is higher and a smaller $EM$ is sufficient to match the emission feature between 6 and 7~keV and consequently the best-fit converges on a slightly smaller value for $N_\mathrm{H}$ and an Fe abundance that is only 5 times solar -- still an order of magnitude more than in 2013.

The photoelectric cross-section of all atoms and ions (in gas and small grains) in the line-of-sight contributes to the X-ray absorption. This is expressed as equivalent hydrogen column density $N_\mathrm{H}$, which is the total hydrogen column density of a gas with a solar abundance pattern. In the energy range between 2 and 3~keV where the observed flux drops (Figure~\ref{fig:spec:rwa}), O, Ne, and Fe are the metals that contribute most to the total photoelectric absorption cross-section \citep{1992ApJ...400..699B}. If these elements are enhanced by e.g.\ one order of magnitude compared to solar abundances, then the true hydrogen column density is one order of magnitude lower than the equivalent hydrogen column density $N_H$ obtained from the fit. The value for $N_\mathrm{H}$ is highly correlated with the temperature and the emission measure. A hotter and less absorbed plasma requires significantly less emission measure to produce the observed flux than a cooler and more absorbed plasma.

Still, the value for $N_\mathrm{H}$ that we fit in either model for the 2017 data should be treated as a lower limit because an absorbing column density this high will hide all signatures of cooler plasma in the spectrum. For example, we could add an extra emission component at 0.6~keV, the temperature seen in 2013, and at the same time increase $N_\mathrm{H}$ without changing the predicted spectrum noticably. In a large survey of T Tauri stars in the Orion nebula cloud, \citet{2005ApJS..160..401P} find that stars with a plasma component of $\mathrm{k}T\approx3$~keV typically also have a cooler component around 1~keV with a comparable $EM$. If that is the case in RW~Aur~A, too, the true  $N_\mathrm{H}$ might be even larger than the value given in table~\ref{tab:specfits} for 2017.

\subsubsection{A detailed look at the emission feature in the RW~Aur~A 2017 spectrum}
\label{sect:adetailedlookattheemissionfeatureintherwaura2017spectrum}
\label{sect:2017}

\begin{figure}[ht!]
    \plotone{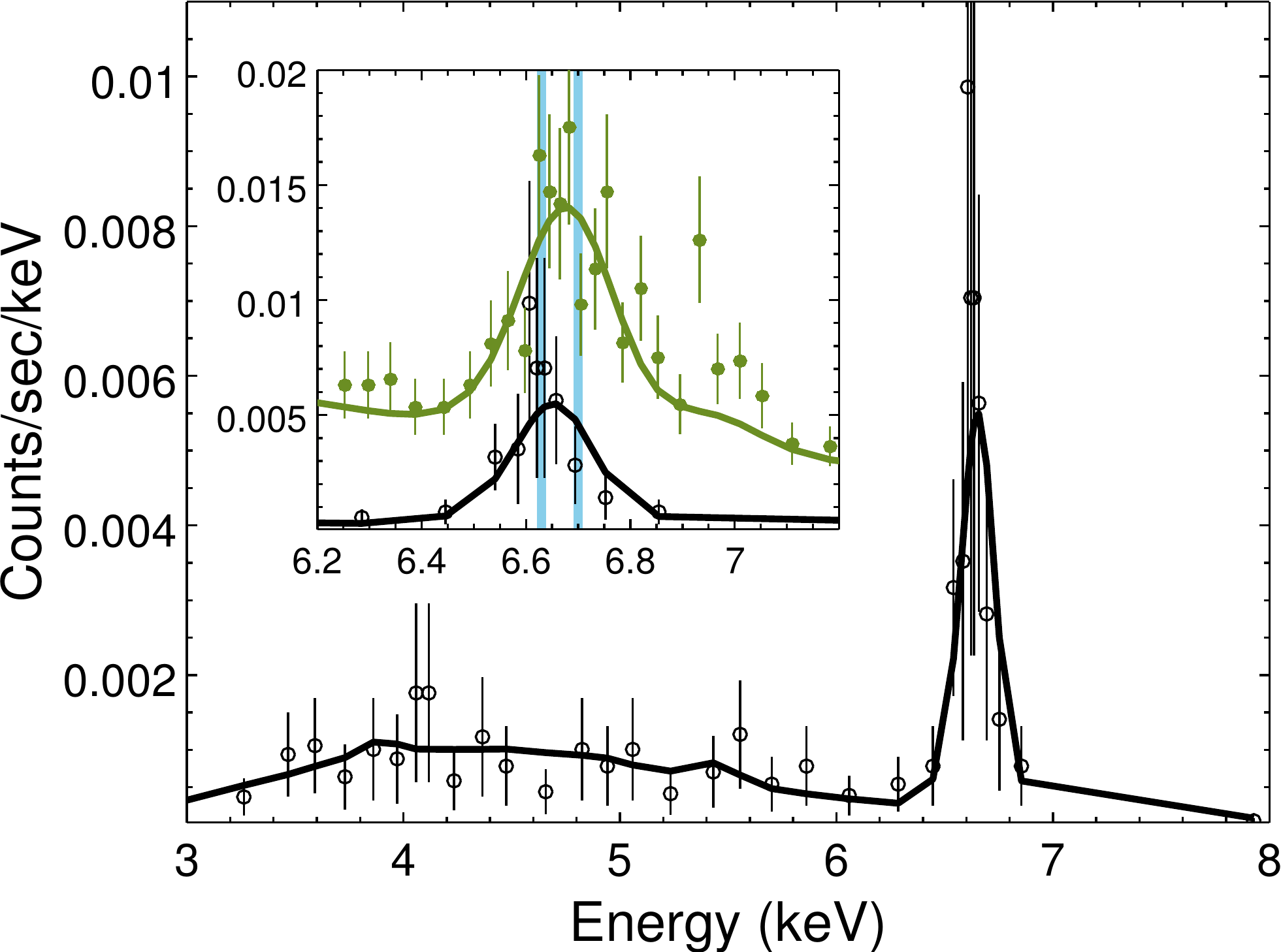}
    \caption{X-ray spectrum for RW~Aur~A observed in 2017 (combined from both observations for presentation purposes) with $1\sigma$ uncertainties and the best fit model overlayed (table~\ref{tab:specfits} (column ``2 temp''). The inset zooms in on the emission feature centered on 6.63~keV. The olive colored line in the inset is data from the flare seen in \emph{XMM-Newton} scaled to the lower effective area of \emph{Chandra}. Due to the emission in a hot flare, the continuum level is much higher and the emission feature is peaked at higher energies than the \emph{Chandra} data. Vertical blue lines in the inset mark the energies 6.63~keV (the centroid of the feature in 2017) and 6.7~keV (the position of Fe~{\sc xxv} emission).}
    \label{fig:spec_17}
\end{figure}
A key feature observed from RW~Aur~A in 2017 is the emission feature between 6 and 7~keV. Figure~\ref{fig:spec_17} shows this spectrum on a linear scale to help locate this feature more precisely. We fit a Gaussian plus a constant to the spectral region 6.0-7.5 keV using the Cash statistic on ungrouped data and find that this feature peaks at $6.63\pm0.02$~keV (90\% confidence interval of statistical uncertainty) $\pm0.02$~keV (99\% confidence interval of the absolute ACIS calibration\footnote{\url{http://cxc.harvard.edu/ccr/proceedings/07_proc/presentations/grant/pg12.html}}). Statistically compatible results are obtained for fits with moderate binning (e.g.\ 5 counts per bin). This region is dominated by a complex of unresolved iron emission lines in the ionization stages Fe~{\sc xxii} to Fe~{\sc xxv}. The He-like triplet of Fe~{\sc xxv} at 6.7~keV dominates the emission if enough hot plasma is present (peak formation temperature for the lines is $\mathrm{k}T=5.5$~keV). The lower ionization stages have weaker lines, many of which are located at slightly lower energies. The peak of the emission feature seen in RW~Aur~A is at 6.63~keV, indicating that the temperature is too low to ionize iron up to Fe~{\sc xxv}. Fe~{\sc xxv} has several lines in the 6.6-6.7~keV range, but the 6.70~keV line is always the strongest one. Thus, if Fe~{\sc xxv} was present, the centroid would be close to 6.70~keV. 
For comparison, an equivalent fit to the \emph{XMM-Newton} data from 2007 where a similar feature is observed finds the peak at $6.68^{+0.03}_{-0.02}$~keV (90\% confidence interval of statistical uncertainty) $\pm0.01$~keV \citep[absolute PN and MOS calibration][]{XMM-SOC-CAL-TN-0018}) which is compatible with Fe~{\sc xxv} emission.

The model for this feature (figure~\ref{fig:spec_17}) is not symmetric, because several unresolved Fe lines from different ionization stages contribute to the emission. The emissivity of this iron feature drops by several orders of magnitude over a small range of temperatures. So, for temperatures at the lower end of the confidence interval, higher iron abundances would be required to match the observed flux at 6.63~keV. That is why the ``2 temp'' model in table~\ref{tab:specfits} requires a higher Fe abundance than the ``1 temp'' model. In the ``2 temp'' model $kT_1$ is lower and thus the emissivitiy of Fe is lower than in the ``1 temp'' model. The position of the observed peak directly rules out a significant contribution from highly-ionized iron, and thus a Fe abundance significantly above solar and at least an order of magnitude higher than in 2013 is needed independent of the choice of parameterization (e.g.\ number of temperature components) for the global model.

In appendix~\ref{appendix:2017} we split the data from 2017 into three phases, but do not find any significant time evolution. In particular, the Fe emission feature is observed at all times; it is not a feature caused by the small flare at the end of ObsID 17764 that can be seen in the lightcurve in figure~\ref{fig:lc}.

\section{Discussion}
\label{sect:discussion}

The optical properties of RW~Aur~AB during the 2016/2017 dimming event are similar to the previous dimming observed from 2014-2016. The $B-V$ and $V-R$ color are similar, and the depth of the dimming is 0.5~mag less than before. In fact, from optical observations alone, it is not clear if the new dimming is an unrelated event or a continuation of the dimming that started in 2014 with a short gap in the absorbing material. Similarly, in X-rays we find that  $N_\mathrm{H}$ increased a factor or 70 or so in 2015 and a few hundred in 2017 (the exact number depends on the plasma model).
On the other hand, in 2017, there is significantly more emission around 5~keV and the Fe abundance is about one order of magnitude higher than in either the optically bright state in 2013 or the optically faint state in 2015.

In this section, we derive limits on the gas and dust mass, distribution, and abundance based on these measurements. In section~\ref{sect:scenarios} we will then discuss which physical processes could cause the changes in gas and dust properties.

\subsection{Where is the absorbing column density located?}
\label{sect:whereistheabsorbingcolumndensitylocated}

For the following discussion it is important to remember that the true hydrogen column density could be lower than the fitted $N_\mathrm{H}$, if elements such as O, Ne, and Fe, which dominate the absorption between 2-3~keV, are enhanced compared to solar abundances.

Using the depth of several optical absorption lines, \citet{2016A&A...596A..38F} placed a limit on the gas phase column density of Na~{\sc i} of $3\times10^{12}\;\mathrm{cm}^{-2} < N_\mathrm{Na} < 2\times10^{14}\;\mathrm{cm}^{-2}$. Using solar photospheric abundances, we can convert this to $1.4\times 10^{18} < N_\mathrm{H} < 9.3 \times 10^{19}\;\mathrm{cm}^{-2}$. This is about one order of magnitude below the $N_\mathrm{H}$ value we observed in 2013 and two to three orders of magnitude below the $N_\mathrm{H}$ we fit in 2015 and 2017 indicating that almost all of the gas in the line-of-sight must be ionized. Na in a low ionization state would be missed in the Na~{\sc i} measurement, but result in same X-ray measured $N_\mathrm{H}$ as neutral Na.

\subsubsection{Are we looking through the disk?}
\label{sect:arewelookingthroughthedisk}

\citet{2006A&A...452..897C} observed the disk of RW~Aur~A in $^{12}\mathrm{CO}$ and $^{13}\mathrm{CO}$ lines with radio interferometry. This allowed them to place limits on the column density of warm gas in the disk of $5\times10^{21}\;\mathrm{cm}^{-2} < N_\mathrm{H} < 10^{23}\;\mathrm{cm}^{-2}$ in the inner 100~au. Their upper limit is close to our measured $N_\mathrm{H}$ and there might be an additional cold gas component in the disk mid-plane that is unseen in the radio observations. So, a sightline through the disk is compatible with our observed value for $N_\mathrm{H}$ in 2017. The problem with this scenario is that the disk has an intermediate inclination. So, the sightline to the central star does not pass through the plane of the disk, unless the inner disk is massively warped. To make matters worse, any sightline though the disk should contain some small dust grains, in contrast to the observed gray absorption (see discussion on the $N_\mathrm{H}/A_V$ ratio in section~\ref{sect:nhav}).

\subsubsection{Are we looking through a disk wind?}
\label{sect:arewelookingthroughadiskwind}

The model fits to the 2015 and 2017 epochs indicate an extremely high absorbing column density of $>10^{23}$~cm$^{-2}$. During the 2014-2016 fading, \citet{2016A&A...596A..38F} and \citet{2016ApJ...820..139T} detected increased blue-shifted Ca~{\sc ii} absorption, indicative of an increasing mass outflow, and they also identify a much stronger [O~{\sc i}]~6300~\AA{} emission line, which is formed in the outflow, compared with spectra taken in the optically bright state. 
\citet{2016MNRAS.463.4459B} argue that this wind also carries the dust that causes the gray absorption although it is not clear if a wind could drag up sufficiently large dust particles from the disk to appear gray \citep{2011MNRAS.411.1104O}. If this wind is strong within the dust evaporation radius, this could provide a large gas column density without any accompanying optical reddening.

As an order of magnitude estimate, we assume that we are looking through a uniform disk wind emanating from the inner 10~au of the disk, and that the line-of-sight does not pass through the disk itself. In order to reach the measured $N_\mathrm{{H}}$ value along the line-of-sight, the density in that region has to be $2 \times 10^9\;\mathrm{cm}^{-3}$.
Even for a modest outflow velocity of only 10~km~s$^{-1}$ this would result in a mass loss rate of more than $2\times10^{-6}\;M_\odot\;\mathrm{yr}^{-1}$. If true, a new knot will become visible in RW~Aur~A's jet very soon.
On the other hand, such mass loss cannot be sustained for very long and would be significantly larger than the mass accretion rate of $4\times10^{-8}\;M_\odot\;\mathrm{yr}^{-1}$ measured in the previous dimming. \citet{2016A&A...596A..38F} even detected a reduced accretion rate in the photometrically dim state, but caution that part of the accretion region could be eclipsed. 

\subsubsection{Are we looking through a screen?}
\label{sect:arewelookingthroughascreen}
A more plausible scenario is that the absorbing column density is not outflowing and the new absorber in 2017 is at least large enough to cover the stellar disk. We can derive a gas mass of
\begin{eqnarray}
m_{\mathrm{screen}} & > & \pi R_*^2 \; l \; n \; 1.3 u  \nonumber \\
  & > & 4 \times 10^{21} \left(\frac{l}{10\;\mathrm{au}}\right) \left(\frac{n}{2 \times 10^9\;\mathrm{cm}^{-3}}\right) \mathrm{ g}   \nonumber \\
  & > & 10^{-6} M_\earth
\end{eqnarray}
where $u$ is the atomic mass unit and $1.3u$ the average particle mass. $l$ is the thickness of the screen and $n$ its particle number density. If the disk is inclined, these masses are easy to explain but they also roughly match the mass of an 80~km size planetesimal, close to the initial planetesimal size \citep{2009Icar..204..558M}. For comparison, this is more than the mass of the Martian moon Phobos \citep[$10^{19}$~g,][]{2014Icar..229...92P} but still five orders of magnitude below the mass of Mercury, the smallest inner planet in our Solar System ($4\times10^{26}$~g).
Similar estimates have been made to explain the added absorption in \object{TWA 30} and \object{T Cha} where \citet{2016MNRAS.459.2097P} and \citet{2009A&A...501.1013S} calculate $5\times10^{19}$~g and $4\times10^{20}$~g, respectively. These calculations only give an order of magnitude estimate for a lower limit to the total mass in this region, since the absorber might be much larger than just the size of the visible star. If the mass is provided by the break-up of a planet or planetesimal, the resulting dust cloud would shear out into a ring and $m_{\mathrm{screen}}$ would represent only a very small fraction of the total mass.  Also, the variability within a dimming event in the optical lightcurve in figure~\ref{fig:lc} suggests that the absorber is not homogeneous.

\subsection{The $N_\mathrm{H}/A_V$ ratio}
\label{sect:then_mathrmha_vratio}
\label{sect:nhav}

\begin{figure}[ht!]
\plotone{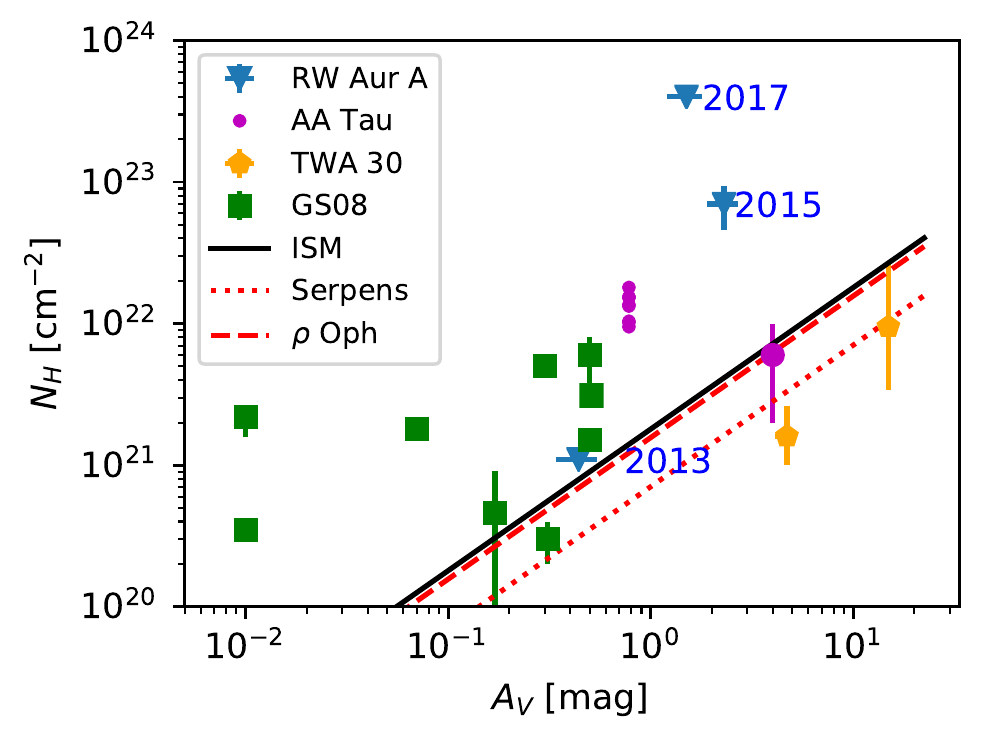}
\caption{$N_\mathrm{H}/A_V$ ratio for RW~Aur~A in the three epochs (see text in section~\ref{sect:nhav} for a description how $A_V$ was estimated for RW~Aur) in comparison to AA~Tau (the small dots without error bars are measurements before the dimming) and TWA~30 as well as the sample from \citet[][GS08]{2008A&A...481..735G}. Also shown is the $N_\mathrm{H}/A_V$ ratio observed in the ISM and the average value for two other star forming regions. See text in section~\ref{sect:nhav} for data sources. Optical and X-ray data for RW~Aur~A, AA~Tau, and TWA~30 are contemporaneous, while the remaining data (GS08, ISM, Serpens, and $\rho$~Oph) are based on optical and X-ray data taken non-comtemporaneously.}
    \label{fig:nhav}
\end{figure}
Figure~\ref{fig:nhav} shows how the $N_\mathrm{H}/A_V$ ratio for RW~Aur~A changes with time. The selective reddening for RW~Aur~A in the optically bright state corresponds to a visual extinction of $A_V = 0.44$ \citep{2015IBVS.6126....1A} and for 2015 and 2017 we add the additional gray extinction seen as $\Delta m_V$ between the long-term average $m_V$ and the $m_V$ in table~\ref{tab:BVR}.
As outlined in section~\ref{sect:introduction}, the additional absorber in the dim phases of RW~Aur~A seems to be gray in the optical photometry; in other words, it is not wavelength selective, because it blocks the light completely. This can either be caused by a screen of large particles such as large dust grains or planetesimals or by a layer of gas and dust with such a high column density that it is optically thick at all relevant wavelengths. In the latter case, however, there has to be a fairly sharp edge between the optically thick absorber, that blocks part of RW~Aur~A and the open space through which we see the remaining flux from RW~Aur~A.
In the X-rays we do not know the true intrinsic source flux. Thus, we cannot reference the previous observations to estimate the true flux, as we do in the optical. It is possible that the gray absorber reduces the overall flux of X-rays in addition to the gas column density $N_\mathrm{H}$ changing the spectral shape.

Figure~\ref{fig:nhav} also shows two other CTTS where $N_\mathrm{H}$ and $A_V$ variability has been seen. AA~Tau used to show variability of $N_\mathrm{H}$ within its 8-day cycle, while the change in $A_V$ over the cycles was not significant \citep{2007A&A...475..607G}. The figure shows these data points as small dots without error bars. Data for AA~Tau's 2013 dimming event are taken from \citet{2015A&A...584A..51S}. TWA~30 also has a time variable $N_\mathrm{H}/A_V$ ratio \citep{2016MNRAS.459.2097P} but is different from RW~Aur~A and AA~Tau as it presents a lower $N_\mathrm{H}/A_V$ ratio than the ISM similar to the Serpens cloud. Moreover, for TWA~30 the values of $N_\mathrm{H}$ and $A_V$ change with time, but the $N_\mathrm{H}/A_V$ could be the same. The two epochs of contemporaneous optical/X-ray observations are separated by more than a month, so the additional absorber can be located well outside of the dust sublimation radius. We see TWA~30 close to edge-on, so the measurements probe the composition of the disk where we expect evolved grains unlike RW~Aur.

$N_\mathrm{H}$ and $A_V$ for the remaining samples in the figure are measured non-simultaneously. 
\citet{2003A&A...408..581V} compared $N_H$ and $A_V$ for six nearby star
forming regions, including the Orion Nebula Cloud (ONC). For the $\rho$~Oph star forming region they
find an $N_H/A_V$ significantly below the ISM value.
They interpret this as a sign that the cloud material has a lower metal abundance than
the ISM, consistent with recent solar abundance measurements.
Alternatively, grain growth can increase the amount of extinction per unit mass
until the grains reach 1 $\mu$m in size \citep{2011A&A...532A..43O}. Grain growths is also the scenario preferred by \citet{2007ApJ...669..493W} to explain the $N_\mathrm{H}/A_V$ ratio seen in the Serpens cloud, where $N_\mathrm{H}/A_V$ is less than half of the ISM value of  $1.8-2.2\times10^{21}\;\mathrm{cm}^{-2}\;\mathrm{mag}^{-1}$ \citep{1995A&A...293..889P}.

\citet{2008A&A...481..735G} analyzed older,
but still accreting classical T Tauri stars (CTTS) in low-mass star forming regions. 
They found consistently
high $N_\mathrm{H}/A_V$ ratios, indicating gas-rich material or massive grain growth. Some
of these sources could be seen through the accretion column located very close to the central star, where small grains evaporate. This is also consistent with observations of AA~Tau where the 8-day periodicity proves that the absorber with high $N_\mathrm{H}/A_V$ ratio is located close to the star, whereas the new absorber, which has a more ISM-like  $N_\mathrm{H}/A_V$ ratio appeared on a much longer time scale and thus must be located at a larger distance from the star outside the dust-sublimation radius.
RW~Aur~A has a fairly large accretion rate so there must be enough mass in the inner disk region, and possibly even within the dust-sublimation radius, to provide a large gas column density. However, there are no indications that the accretion rate changes significantly between the dim state and the bright state \citep{2016A&A...596A..38F}. Thus, the increased $N_\mathrm{H}$ points to a change in the geometry rather than a change in the accretion rate.

\subsection{The extra emission measure}
\label{sect:theextraemissionmeasure}

The \emph{Chandra} X-ray and optical lightcurves of RW~Aur~A (figure~\ref{fig:lc}) in 2017 are mostly flat and the fact that a similar flux level was observed two days later shows that the observations were not taken during a big flare. Thus, the increased emission at energies above 3~keV in 2017 compared to 2013 and 2015 (figure~\ref{fig:spec:rwa}) must be due to a different structure of the emission region.

The energy of emission from shock heated material is limited by the velocity jump across the shock front. For the mass and radius of RW~Aur~A \citep{2001ApJ...556..265W,2012AstL...38..167D}, the free fall velocity is $<500\;\mathrm{km s}^{-1}$, and thus the temperature of the shock $<0.3$~keV. The relatively soft X-rays observed by \citet{2014ApJ...788..101S} in the resolved jet show that shocks in the outflow reach similar temperatures. Since the temperature of the observed plasma in 2017 is considerable larger, it must be magnetically heated in the corona.

\subsection{Fluorescent iron?}
\label{sect:fluorescentiron}

When iron is ionized by high-energy photons, an inner electron from the K-shell can be removed. Fluorescence occurs when an electron from a higher level (usually L) recombines to fill the K-shell. The energy of the fluorescent emission depends on the charge state of the ion. For neutral iron or ions in a low ionization stage the energy is around 6.4~keV. Only for highly ionized Fe does this line reach an energy similar to the feature we observe in the 2017 observation of RW~Aur~A \citep[Fe~{\sc xxiii}: 6.63~keV,][]{2004ApJS..155..675K}. Fluorescent Fe emission is sometimes seen in CTTS when very bright flares occur and provide a high X-ray flux that ionizes neutral iron in the accretion disk or funnels \citep[e.g.][]{2005ApJS..160..503T,2007A&A...470L..13C,2010ApJ...714L..16H}. In RW~Aur~A, we see a feature centered on 6.63~keV. If this is a fluorescence line, the source of it is so highly ionized that it must be located in the stellar corona, while at the same time, the source of the ionizing radiation must be hidden from view, since we do not detect a 6.7~keV feature. Thermal emission from a range of Fe species (mostly Fe~{\sc xxii}-Fe~{\sc xxiv}) as discussed in Section~\ref{sect:2017} is a simpler and more likely explanation for the observed spectrum.

\subsection{The Fe abundance}
\label{sect:thefeabundance}

The best model fit to the X-ray spectrum of RW Aur A in 2017 in table~\ref{tab:specfits} (model ``2 temp'') shows an Fe abundance compared to solar of $\sim 15$ and an emission measure of $50\times10^{52}$~cm$^{-3}$.
An increase in iron abundance is not compatible with the properties of an active stellar corona. There is some element differentiation in coronae, where elements of low FIP such as Fe are enhanced in stars with low activity and depleted in stars with a high activity level \citep[see review by][and references therein]{2004A&ARv..12...71G}. RW~Aur~A had a low Fe abundance in 2013, which is compatible with this picture. However, in 2017 there is significantly more emission at high temperature, indicating a higher level of activity and thus the Fe abundance should have decreased instead of increased.
We use the $EM$ and abundance to estimate the total mass of iron in the emitting material, which is inversely proportional to the assumed density. To simplify the estimate, we assume that the emitting volume $V$ is filled by plasma with a constant ion number density $n$ and an electron density $n=1.4\;n_e$, where $1.4$ is the average number of electrons released per ion. We can then write $EM = V\;n\;n_e=1.4\;V\;n^2$. We calculate the total mass of Fe by multiplying the total number of Fe ions with the average weight of an Fe ion ($55.8\;u$), where $u$  the atomic mass unit.
\begin{equation}
m_{\mathrm{Fe}} = V\;n\;a\;55.8\;u
\end{equation}
Assuming a typical coronal density of $n=10^{10}\;\mathrm{cm}^{-3}$ \citep{2004A&A...427..667N}, the reference Fe abundance from \citet{2009ARA&A..47..481A} and the values for $EM$ and relative Fe abundance fitted in table~\ref{tab:specfits} (2017, column ``2 temp'') only about $3 \times 10^{-10}$ Earth masses of iron are required in the emitting plasma. Since the plasma temperature rules out an origin in the accretion shock, any accreted mass must be transferred to the corona in some way. Observations of RS~CVn EI~Eri suggest that the time scale of element fractionation in active regions is a few days \citep{2013A&A...550A..22N}; similarly, the time scale on which mass lost to the solar wind is replaced is one to two days \citep{2015LRSP...12....2L}. Assuming that one day is also a reasonable estimate for the time that accreted iron remains in the corona before it is mixed in the convective zone or ejected into the stellar wind and assuming that the abundance we observe is typical for the most recent dimming starting in 2016, about $10^{-7}$ earth masses of iron have passed through the corona. A planetesimal could easily supply this reservoir. At an age $<10$~Myr \citep{2001ApJ...556..265W} the convection zone of RW~Aur~A is still so deep that it contains about half of the total stellar mass \citep{2011ApJ...743...24S}; the convective turnover time is of order one year \citep{2010A&A...510A..46L}. Thus, Fe would accumulate in the upper photosphere until it reaches a maximal concentration after about one year.

\section{Scenarios}
\label{sect:scenarios}

We are looking for a unified model that explains the variable absorption, Fe abundance, and increased volume of magnetically heated plasma.
The most obvious explanation for additional flux at high energies in an active star is a large coronal flare \citep[e.g.][]{2005ApJS..160..469F}. However, the X-ray and optical light curves during the observations in 2017 show no indication of a flare. Also, coronal flares induce little change in $N_\mathrm{H}$, in contrast to our observations.
In the following sub-sections we discuss other scenarios; a planetesimal break-up or puffing up the inner disk seem to be most consistent with the data.

\subsection{Can this be explained by the tidal stream?}
\label{sect:canthisbeexplainedbythetidalstream}

\citet{2013AJ....146..112R} suggested that the dimming event in 2011 could be due to the tidal stream between RW~Aur~A and B passing through the line-of-sight. However, many of the features observed in the more recent dimming events can only be explained by variable phenomena near the location of the inner disk and/or the disk wind close to the star \citep[e.g.][]{2015IBVS.6143....1S,2015A&A...577A..73P,2016ApJ...820..139T,2016A&A...596A..38F,2016MNRAS.463.4459B} which is consistent with the data presented here: A stream of gas and dust passing by at a large distance from the star cannot cause the changes in the emission region needed to explain the increased hot emission and the Fe abundance inferred from X-ray spectral modeling (section~\ref{sect:2017}). Strictly speaking, this does not rule out that the 2011 event was caused by the tidal stream passing through our line-of-sight and any later dimming is due to an unrelated mechanism, but given the long-term stability of the lightcurve before 2011 \citep{2013AJ....146..112R} it seems more likely that all dimming events are related in some way.

\subsection{Planetesimal break-up in the inner disk}
\label{sect:planetesimalbreakupintheinnerdisk}

Circumstellar disks are the sites of planet formation. As part of this process, dust grains coagluate into larger aggregates and eventually into planetesimals and planets. When two particles collide, they may either stick together or break apart. Typical disk lifetimes are a only few Myrs \citep[see review by][]{2014prpl.conf..475A}, but RW~Aur~A still has a disk at an apparent age of 10~Myr \citep{2001A&A...376..982W}, so the system certainly had enough time to form planets, and given the long time scale, it may have formed more planets or planetesimals than typical.

A possible scenario is that two large planetesimals collided in 2011 and released a cloud of smaller particles, which caused the optical dimming. After about 6 months, the particles are no longer visible because they are accreted onto the star or settled into the disk midplane.

However, some larger fragments of the collision may remain and the collision may have set them on eccentric orbits increasing the probability of further collisions after the initial event. We speculate that collisions caused the dimmings in 2014 and possibly again at the end of 2016. The products of each collision depend on the composition of the colliding planetesimals and the impact parameter \citep{2012A&A...540A..73W}. Collision products can collide again and cause a cascade that break up particles down to mircometers \citep{2014A&A...566L...2K}. If the size distribution is skewed to particles that are more than $\mu$m sized this will cause gray absorption in the optical. At the same time, the newly released dust will increase the opacity of the disk in layers that were optically thin before. This leads to more energy absorption and rising temperatures, which will increase the scale height of the disk, pushing the limits of optically thick absorption to even larger heights. Detailed modeling is needed, but this is at least consistent with the observed increased optical and X-ray column density.

We can estimate the dust mass in the line of sight assuming spherical dust grains with radius $a$ and mass density $\rho=1\mathrm{g~cm}^{-3}$ distributed over a length $l$ with a number density $n$. They will reduce the flux by a factor $e^{-\pi a^2 n l}$. The optical light curve shows a dimming of $\Delta V=2-3$~mag, about a factor of 10. As a minimum, the dust column covers the stellar disk, so a lower limit to the dust mass $M_d$ of the gray absorber is
\begin{equation}
M_{\mathrm{dust}} > \pi R^2_* \frac{\ln{10}}{\pi a^2}\;\frac{4}{3}\pi a^3 \;\rho \approx 5\times10^{19}
\;\mathrm{g}\; \left(\frac{a}{10\;\mu\mathrm{m}} \right) \ .
\end{equation}

For grains with $a=0.1\;$mm, the dust mass is 1/100 of the gas mass estimated from the $N_\mathrm{H}$ in table~\ref{tab:specfits} (column 2017, ``2 temp'') assuming solar abundances in the absorbing material, i.e.\ a mean particle mass of $1.3\;u$. This ratio is similar to common ISM values. However, for a particle size distribution $n(a)\propto a^{-3.5}$, $a$ in the equation above is replaced by $\sqrt{a_0 a_1}$, where $a_0$ is the smallest particle (we take 0.1~mm) and $a_1$ is the largest particle. Both $a_0$ and $a_1$ are not well known, but $a_1 = 1$~km is a reasonable scale for the break up of a planetesimal, which would lead to an $M_{\mathrm{dust}}$ that matches the gas mass from $N_H$ within a factor of a few.

The X-ray spectrum taken in 2015 did not show any peculiar Fe feature and emission in general was fainter than in 2017. Either accretion of Fe is intermittent, or the collision that caused the 2017 dimming event happened to include a more Fe rich planetesimal than the previous events. RW~Aur~A is old enough that considerable planet migration may have happened \citep[e.g.][]{2002ApJ...565.1257T} and thus the inner disk region may contain planets or planetesimals that formed in different regions of the disk and thus have different compositions.

\subsection{Context for a planetesimal break-up scenario}
\label{sect:contextforaplanetesimalbreakupscenario}

Detailed hydrodynamical modeling of disk and planetesimal for the scenario discussed here is beyond the scope of this paper, but we can compare the time scale of the changes in the lightcurve with the time scale of the relevant physical processes in the disk to constrain the location of the event. The optical dimming observed in 2011 lasted six months. Another dimming event started in 2015 and lasted until at least the end of 2017 with just short spikes in the lightcurve up to the bright state level. X-ray observations have been done about every two years and the plasma properties change significantly in between. 

Assuming that the first break-up is due to a collision around 2011, it could have set fragments onto eccentric trajectories. For RW~Aur~A, the Roche limit is about $2\;R_*$, which is well inside the co-rotation radius. In the magnetically funneled accretion scenario, the stellar magnetic field couples to the disk around the co-rotation radius and little gas or particles exist inside for more than about one orbit. Thus, we expect that the planetesimal break apart due to collisions, not tidal interaction with the central star. A collision in 2011 and 2015, which releases a cloud of gas and dust might cause the dimming and increase in $N_\mathrm{H}$ from 2015 on. Can this also explain the sudden increase of the Fe abundance? In a circumstellar disk, the gas rotates at slightly sub-keplerian velocities, because it is supported by gas pressure. On the other hand, dust particles need to rotate at keplerian velocities to maintain a stable orbit, thus they feel a headwind and migrate inwards; the rate of this migration depends on their Stokes number. The radial drift time for particles is a few hundred orbits, so drifting from a radius just slightly outside the co-rotation radius to the inner edge of the disk (where gas and particularly Fe-rich particles couple to the magnetic field and are accreted) could happen in about 1.5 years \citep{2016SSRv..205...41B} for particles with a Stokes number around 1 \citep{2008A&A...480..859B}. This corresponds to a radius of $a = 2 \frac{\Sigma_g}{\pi\rho} = 54$~m where $\Sigma_g$ is the gas surface density of the disk and we have used $\Sigma_g=200\;\frac{\mathrm{g}}{\mathrm{cm}^2}\left(\frac{r}{\mathrm{au}}\right)^{-1}$ at disk radius $r$. 
The value of $\Sigma$ close to the star is not well known. $\Sigma_g=200\;\frac{\mathrm{g}}{\mathrm{cm}^2}$ at 1~au is lower than typically estimated values for T~Tauri disks \citep{2010ApJ...723.1241A} and higher than expected values for the disk mass estimates in the RW~Aur~A simulations of \citet{2015MNRAS.449.1996D}.

Under these assumptions, the first particles from the 2015 fragmentation just made it onto the central star in 2017. It is not unreasonable, that those fragments originated in the compact core of the original planetesimal and are Fe rich. We should see enhanced accretion in the coming years as both larger and smaller fragments pass through the disk and are accreted. 
\citet{2018arXiv180409190R} point out earlier dimming events in 1937 and 1987, so collisions of this type might be common in the RW~Aur~A disk due to the disturbance from RW~AurB.

A scenario that essentially requires the accretion of a planetesimal or (part of) a terrestrial planet is not without precedent. While this has never been observed directly in young stars, there are several white dwarfs whose surface abundances show clear signatures of ongoing accretion of debris from a terrestrial planet \citep{2006ApJ...653..613J,2009ApJ...694..805F,2011ApJ...732...90M,2012MNRAS.424..333G}.

\subsection{Other scenarios}
\label{sect:otherscenarios}

The precession time scale for a disk similar to RW~Aur should be about 0.5~Myr at 40~au \citep{2017MNRAS.469.2834O}, which is 2800 orbits. Even when scaling this down to the co-rotation radius, that still predicts a scale of 80 years; furthermore, such precession should be periodic, but RW~Aur~A's lightcurve before 2011 did not show disk eclipses for at least one century. 

There is however a possible scenario that shares many characteristics with a planetesimal break-up: a pressure trap. Thick disks can contain deadzones where the ionization is so low that the magnetic field does not couple to the disk material. \citet{2016A&A...596A..81P} show that particles with $>1$~mm will be trapped on the outside of a deadzone. If the disk structure changes, e.g.\ responding to an inwards traveling wave excited by the tidal forces of RW~Aur~B passing by \citep{2015MNRAS.449.1996D}, and that trapped material is released, this might supply enough large grains to explain the extra absorber. As in the planetesimal collision scenario, particles of a certain size would move inward and supply the Fe seen in the X-ray emission. If this is true, the accretion rate in RW~Aur~A should increase significantly in the next few years as the bulk of the mass reaches the inner disk edge.

Last, we note that radial velocities of several 1000~km~s$^{-1}$ would be required to explain the 6.63~keV feature as a red-shifted 6.7~keV Fe~{\sc xxv} line. This is an order of magnitude faster than the fastest known jet component in RW~Aur. \citet{2014ApJ...788..101S} observed X-ray emission from the jet, which requires just a few hundred km~s$^{-1}$ for shock-heating \citep[e.g.][]{2014ApJ...795...51G}.

\section{Summary}
\label{sect:summary}
We present new \emph{Chandra} data of the binary RW~Aur. The resolved binary member RW~Aur~A had several optical dimming events between 2011 and 2017. Previously published \emph{Chandra} data show RW~Aur~A in an optically bright state and in a previous dimming event. We find that RW~Aur~A has an exceptionally high absorbing column density of a few $10^{23}\;\mathrm{cm}^{-2}$ in 2017, orders of magnitude more than in the optically bright state. We also see significantly enhanced emission in the hard X-ray range above 3~keV and newly appeared Fe emission feature at 6.63~keV that indicates an Fe abundance one order of magnitude above solar. The temperature of the plasma is too high to be shock-heated; the most plausible location for it is an active corona. Significant accretion of Fe rich material is required to boost the abundance to the observed value. We speculate that the break-up of a terrestrial planet or a large planetesimal might supply the gray extinction seen in the optical, the large amount of gas column density observed as $N_\mathrm{H}$ in X-rays and also provide the iron in the accretion stream to enhance coronal abundances.

\acknowledgments
We thank an extremely thorough and detail oriented referee for numerous suggestions that lead to substantial improvements in the manuscript.
The scientific results reported in this article are based on observations made by the Chandra X-ray Observatory. We acknowledge with thanks the variable star observations from the AAVSO International Database contributed by observers worldwide and in particular the BAAVSS. This research has made use of software provided by the Chandra X-ray Center (CXC) in the application packages CIAO, ChIPS, and Sherpa. Support for this work was provided by the National Aeronautics and Space Administration through Chandra Award Numbers DD5-16077X and GO6-17021X issued by the Chandra X-ray Observatory Center, which is operated by the Smithsonian Astrophysical Observatory for and on behalf of the National Aeronautics Space Administration under contract NAS8-03060.
TB acknowledges funding from the European Research Council (ERC) under the European Union's
Horizon 2020 research and innovation programme under grant agreement No 714769.
PCS acknowledges support by DLR grant 50 OR 1706.
\vspace{5mm}
\facilities{CXO(ACIS,ACA), AAVSO}


\software{astropy \citep{2013A&A...558A..33A},
          CIAO \citep{2006SPIE.6270E..60F},
          Sherpa \citep{2007ASPC..376..543D}
          }

\appendix

\section{Time resolved analysis of the RW Aur A \emph{Chandra} data from 2017}
\label{sect:timeresolvedanalysisoftherwauraemphchandradatafrom2017}
\label{appendix:2017}

Figure~\ref{fig:lc} shows that the count rate in RW~Aur~A increases towards the end of the first observation taken in 2017; this might be due to a coronal flare. A natural approach is to search for time-variability in the X-ray spectrum, in particular with an eye towards the question if the Fe emission feature is correlated with that flare. This appendix complements the analysis in sections~\ref{sect:chan:rwa} and \ref{sect:2017}.

We split the longer of the two observations (ObsID 177654) in 2017 into two parts, a ``pre-flare'' block (up to 30~ks) and a ``flare'' block (after 30~ks). We then fit those two spectra and the spectrum from ObsID~19980, which was taken two days later, with a similar model to the one used in section~\ref{sect:chan:rwa}. Since the count rate in each of the three spectra is low we use the Cash statistic on ungrouped data and we only present fits for a model with a single temperature component. We use the same background model as in section~\ref{sect:chan:rwa}, i.e.\ we describe the contamination from RW~Aur~B with the model fit for 2017 in table~\ref{tab:specfits} and a single scale factor to account for the fact that only a small fraction of the RW~Aur~B PSF falls into the extraction region of the RW~Aur~A spectrum. RW~Aur~B is moderately variable in 2017, but the variability appears not to be correlated with RW~Aur~A. Furthermore, the signal from RW~Aur~A is so low that the uncertainties of the fitted parameters are much larger than any change that would be introduced by relatively small changes in the background model.

\begin{figure}[ht!]
    \plotone{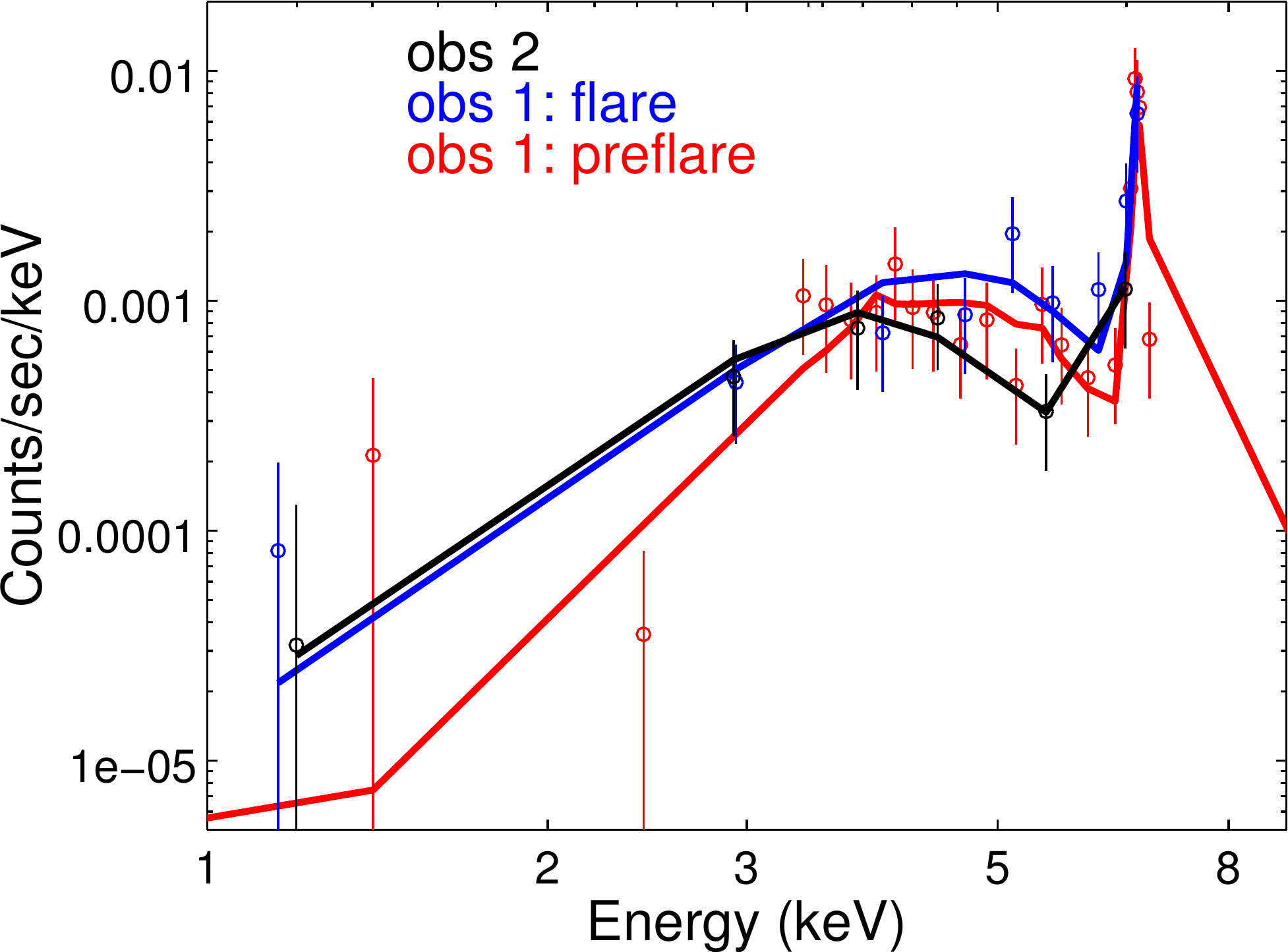}
    \caption{X-ray spectra for RW~Aur~A in 2017. Points are observational data and lines show models from table~\ref{tab:specfits17}. The first observation in 2017 (ObsID 17764) is split into a ``preflare'' state (the first 30~ks of the exposure) and a ``flare'' state (the remaining exposure). The second observation is ObsID 19980. Observations are binned to 5 counts per bin and background-subtracted for display purposes. Error bars indicate $1\sigma$ uncertainties.}
    \label{fig:spec:2017_flare}
\end{figure}

Figure~\ref{fig:spec:2017_flare} shows the observed data and the best-fit models. Model parameters are listed in table~\ref{tab:specfits17}. The signal is weak in all three spectra. The figure shows marginally stronger emission in the 3-5~keV range in the ``flare'' spectrum  than in the other two spectra, which is consistent with the interpretation that the increase in flux in the lightcurve towards the end of ObsID~17764 is related to stellar activity. Figure~\ref{fig:spec:2017_flare} indicates that the Fe emission feature at 6.63~keV is present in all spectra, but the low signal makes this hard to see in the binned data shown in the figure.

The best-fit values vary between the three spectra but uncertainties are so large that all three fits are still compatible with each other. The numbers highlight some of the ambiguities in the fits. For example, while the best-fit k$T$ increases in the flare, the derived intrinsic $L_X$ in the flare is actually lower than the pre-flare $L_X$, because the fitted pre-flare $N_{\mathrm{H}}$ is also larger. Similarly, we can see the strong dependence of the Fe abundance on the fitted temperature. The fit for the ``flare'' phase has such a large best-fit temperature that Fe is so highly ionized that little Fe~{\sc xxv} remains, thus a very large Fe abundance is needed to explain the emission feature. On the other hand, the fit for ObsID 19980 is cool enough to have Fe in the appropriate ionization stages and consequently the fit gives a lower Fe abundance. Again, the values are compatible within the uncertainties. Some of this is due to our simple model with a single temperature emission component, while in real coronae we will always see a temperature distribution. However, the signal is too low to make a fit with several temperature components meaningful.
It is noteworthy though, that all scenarios show a Fe abundance significantly above solar.

\begin{table}
\caption{\label{tab:specfits17} Parameters of X-ray model fits for RW Aur A in 2017. Uncertainties are 90\% confidence ranges.}
\begin{tabular}{lcccc}
\hline \hline
ObsID & combined & 17764 & 17764 & 19980\\ 
      & ``1 temp'' from table~\ref{tab:specfits} & quiescent & flare & obs 2 \\
\hline
$N_\mathrm{H}^a$ [$10^{21}$~cm$^{-2}$] & $300 \pm 100$ & $400^{+300}_{-150}$ & $200_{-100}^{+300}$ & $210_{-150}^{+170}$ \\
k$T_1$  [keV]& $3.1^{+1.8}_{-1.3}$ & $2.3^{+2}_{-1.3}$ & $13^{+8}_{-10}$ & $1.6^{+18}_{-0.6}$\\
$EM_1$  [$10^{52}$~cm$^{-3}$] & $17^{+45}_{-9}$ & $35_{-25}^{+300}$ & $0.7_{-0.6}^{+17}$ & $17_{-17}^{+94}$\\
abund$^b$: Fe & $5^{+3}_{-2}$ & $4_{-2}^{+1000}$ & $454_{-450}^{+4000}$ & $10_{-8}^{+4000}$ \\
abund$^b$: Ne & =1 & =1 & =1 & =1\\
\hline
statistic & \multicolumn{4}{c}{Cash}\\
\hline
observed flux$^c$ & 2.9 & 2.6 & 8.7 & 1.6\\
intrinsic $\log L_X^d$ & 30.6 & 30.8 & 30.5 & 30.9\\
\hline
\end{tabular}\\
$^a$: These numbers should be treated as lower limits. See section~\ref{sect:chan:rwa} for discussion.\\
$^b$: Relative to solar abundances from \protect{\citet{2009ARA&A..47..481A}}\\
$^c$: in units of $10^{-13}$erg~s$^{-1}$~cm$^{-2}$ for the energy range 0.3-9.0~keV. Value is given for best fit model without uncertainties.\\
$^d$: in units of erg~s$^{-1}$ for the energy range 0.3-9.0~keV. Value is given for best fit model without uncertainties.
\end{table}
\begin{figure}[ht!]
    \plotone{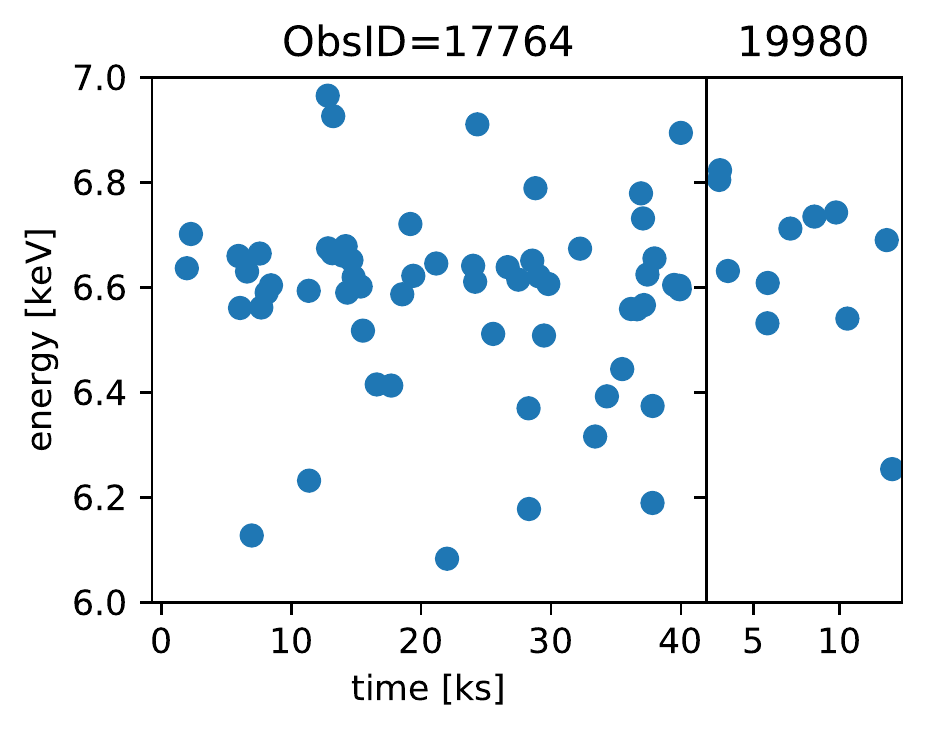}
    \caption{Arrival times of individual photons in the Fe feature for the two observations taken in 2017. Because of the limited energy resolution of the detector we cannot say for an individual photon if it is emitted in the continuum or in a line but the figure shows that photons between 6.5 and 6.8~keV, which are likely to come from the Fe emission feature, are detected throughout the observation.}
    \label{fig:17timing}
\end{figure}
Figure~\ref{fig:17timing} shows the individual photon arrival times for the observations in 2017. Since the energy resolution of the detectors is limited, energies for photons in the emission feature scatter around 6.63~keV. In addition there is a weak continuum. The figure shows that photons in the 6.63~keV feature are detected through both observations and that they are not clustered towards the end of ObsID 17764, when the lightcurve rises. The emission feature seems not to be associated with flare emission, but is consistently seen in the quiet corona.

\bibliographystyle{../AAStex/v611/aasjournal}
\bibliography{../articles}

\end{document}